\documentclass[journal=apchd5,manuscript=article,layout=twocolumn]{achemso}

\usepackage[version=3]{mhchem} % Formula subscripts using \ce{}
\usepackage{dsfont}
\usepackage{mathtext}
\usepackage{mathtools}
\usepackage{amsmath}
\usepackage{amssymb,mathrsfs}
\usepackage{graphicx}
\usepackage{bm}
\usepackage{textcomp}
\usepackage{appendix}
\usepackage{color}

\newcommand{\bi}{{\bm i}}
\newcommand{\bj}{{\bm j}}
\newcommand{\br}{{\bm r}}

\newcommand{\tphi}{\tilde{\phi}} 
\newcommand{\tpsi}{\tilde{\psi}} 
\newcommand{\tH}{\tilde{H}}

\let\oldmaketitle\maketitle
\let\maketitle\relax

\makeatletter
\let\l@addto@macro\relax
\makeatother
\usepackage[fontsize=10pt]{scrextend}

\title{Vector Topological Edge Solitons in Floquet Insulators}

\author{Sergey K. Ivanov}
\affiliation{Moscow Institute of Physics and Technology, Institutsky lane 9, Dolgoprudny, Moscow region, 141700, Russia}
\alsoaffiliation{Institute of Spectroscopy, Russian Academy of Sciences, Fizicheskaya Str., 5, Troitsk, Moscow, 108840, Russia}
\email{sergey.ivanov@phystech.edu}

\author{Yaroslav V. Kartashov}
\affiliation{Institute of Spectroscopy, Russian Academy of Sciences, Fizicheskaya Str., 5, Troitsk, Moscow, 108840, Russia}
\alsoaffiliation{ICFO-Institut de Ciencies Fotoniques, The Barcelona Institute of Science and Technology, 08860 Castelldefels (Barcelona), Spain}

\author{Alexander Szameit}
\affiliation{Institute for Physics, University of Rostock, Albert-Einstein-Str. 23, 18059 Rostock, Germany}

\author{Lluis Torner}
\affiliation{ICFO-Institut de Ciencies Fotoniques, The Barcelona Institute of Science and Technology, 08860 Castelldefels (Barcelona), Spain}

\author{Vladimir V. Konotop}
\affiliation{Departamento de F\'isica and Centro de F\'isica Te\'orica e Computacional, Faculdade de Ci\^encias,
Universidade de Lisboa, Campo Grande, Ed. C8, Lisboa 1749-016, Portugal}


\begin{document}

%%%%%%%%%%%%%%%%%%%%%%%%%%%%%%%%%%%%%%%%%%%%%%%%%%%%%%%%%%%%%%%%%

\twocolumn[
\begin{@twocolumnfalse}
\oldmaketitle
\begin{abstract}
We introduce topological vector edge solitons in a Floquet insulator, consisting of two honeycomb arrays of helical waveguides with opposite directions of rotation in a focusing nonlinear optical medium. Zigzag edges of two arrays placed in contact create a zigzag-zigzag interface between two structures with different topology. A characteristic feature of such a photonic insulator is that, in the linear limit, it simultaneously supports two topologically protected chiral edge states at the interface between the two arrays. In the presence of nonlinearity, either bright or dark scalar Floquet edge soliton can bifurcate from a linear topological edge state. Such solitons are unidirectional and are localized in both directions, along the interface due to nonlinear self-action, and across the interface as being an edge state. The presence of two edge states with equal averaged group velocities enables the existence of stable topological vector edge solitons. In our case these are nonlinearly coupled bright and dark solitons bifurcating from different branches of the topological Floquet edge states. Here we put forward a new mathematical description of scalar and vector small-amplitude Floquet envelope solitons in the above-mentioned continuous system. Importantly for the design of future photonic devices based on Floquet edge solitons, we find that the latter can be described by nonlinear Schr\"odinger equations for the mode envelopes obtained by averaging over one rotation period in the evolution coordinate.
\end{abstract}
\end{@twocolumnfalse}
]

\small
%%%%%%%%%%%%%%%%%%%%%%%%%%%%%%%%%%%%%%%%%%%%%%%%%%%%%%%%%%%%%%%%%

\section*{Introduction}

Topological insulators, originally introduced and still very actively studied in solid state systems (see reviews \cite{elect1,elect2} and references therein), feature the unique ability to support topologically protected edge states enabling conductance, even though in the bulk the materials behave as conventional insulators. Nowadays topological insulation is recognized as a universal physical phenomenon encountered in diverse areas of science. Topological insulators and topologically protected transport were introduced theoretically and demonstrated experimentally in mechanical systems \cite{mech1}, acoustics \cite{acou1,acou2}, systems of cold atoms in optical lattices \cite{cold1,cold2}, atomic Bose-Einstein condensates \cite{cond1}, polariton condensates \cite{polar1,polar2}, and, most notably, in optical systems \cite{Haldane} (see also reviews \cite{topphot1,topphot2}), including gyromagnetic photonic crystals \cite{giro1,giro2}, arrays of coupled resonators \cite{res1,res2}, arrays of helical waveguides \cite{helix1}, metamaterial superlattices \cite{meta1}, and various dissipative systems allowing realization of topological lasers \cite{laser1,laser2,laser3,laser4}.

One of the most successful approaches to the realization of topological insulators in optical systems relies on periodic driving, i.e., on periodic modulation of the parameters of a system in the evolution variable. The importance of periodic driving as a powerful tool for control and manipulation of the evolution of the excitations was recognized long ago in quantum mechanics (see e.g. the reviews~\cite{pdrive1,pdrive2}, as well as more recent studies~\cite{Kitagawa-10,Zhenghao-11,well1,pdrive3}). Periodically driven optical or optoelectronic systems have been used for the implementation of anomalous topological insulators \cite{helix2,helix3,cones1}; the observation of optical Weyl points \cite{Noh-17}, topological Anderson insulators \cite{anderson1,anderson2,anderson3}, topologically protected bulk states in synthetic dimensions \cite{Lustig-19}, topologically protected path entanglement \cite{Rechtsman-16}, fractal topological spectrum in helical quasicrystals \cite{Bandres-16}, guiding light by the artificial gauge fields \cite{Lumer-19}, resonant switching \cite{respol1} and control of the propagation velocity \cite{respol2} of the topological states, and several other spectacular phenomena. It was also shown that driven topological systems, such as helical arrays, can be characterized by special topological invariants~\cite{Rudner-13}.

Periodic driving may break time-reversal symmetry. In combination with Dirac degeneracies in the spectrum of the system, this may lead to the appearance of unidirectional topologically protected edge states, which are immune to backscattering~\cite{Kitagawa-10,well1,Zhenghao-11}. In Optics this phenomenon was observed experimentally in helical waveguide arrays~\cite{helix1}.

One of the major advantages of the optical systems, that remains largely unexplored in the context of the Floquet insulators, is that such systems can be strongly nonlinear. Nonlinearity opens new routes for the control of the topological edge states, and, most importantly, it enables new phenomena. These include modulational instability and topological edge solitons, which inherit topological protection against scattering into the bulk modes and remain localized upon motion along the edge of a topological insulator. To date, topological edge solitons have been  studied mainly in undriven systems, such as polariton topological insulators \cite{polsol1} admitting bright \cite{polsol2} and dark \cite{polsol3,polsol4} quasi-solitons, as well as rich bistability effects~\cite{polsol5}. 

Solitons in Floquet insulators have received less attention. So far, only compact scalar (single-component) self-sustained excitations have been found numerically at the self-induced edges~\cite{bulkFsol} or at real interfaces \cite{edgeFsol1,edgeFsol2} of the nonlinear Floquet topological insulators. Such states emerge upon the development of the modulational instability of the edge states. While tight-binding description of the edge solitons in Floquet topological insulators has been developed in \cite{discrete1,discrete2}, the theory of such unusual excitations describing their bifurcation from the linear Floquet-Bloch modes for genuinely continuous systems, in the limit opposite to the tight-binding one, is still missing. Indeed, the development of such a description is not straightforward in modulated systems, because the model should describe bifurcation of the families of the envelope solitons from the continuous linear Floquet-Bloch states performing regular $z$-dependent oscillations, i.e., the description of the evolution of the $z$-averaged envelope must be constructed.

The aim of this paper is twofold. First, we present a mathematical description of the envelope topological edge solitons  existing in periodically driven continuous optical systems that substantially differs from the standard theory of gap solitons in stationary (i.e., undriven) periodic systems (see e.g.~\cite{KS,solit-in-latt}). We show that the developed theory accurately describes the profiles of the envelope Floquet edge solitons in continuous arrays of helical waveguides and allows to determine optimal for their excitation parameters of the underlying helical structure, thereby laying the ground for the experimental observation of such states. Second, we propose a realistic structure involving two waveguide arrays with opposite directions of waveguides rotation that can support two coexisting edge states at one interface. We show that solitons constructed on these two states can couple to form previously unknown dark-bright composite vector Floquet edge solitons, and extend our theory to this vector system.

%%%%%%%%%%%%%%%%%%%%%%%%%%%%%%%%%%%%%%%%%%%%%%%%%%%%%%%%%%%%%%%%%
\section*{Theory of envelope solitons in Floquet insulators}
\label{sec:theory}

\subsection*{Floquet-Bloch states}
\label{subsec:FB}

We address the propagation of light in a Kerr nonlinear medium whose refractive index is modulated periodically along the propagation axis $z$ and in the transverse plane along $y$ direction. Such medium is described by a periodic optical potential $U(\br,z)=U(\br,z+T)=U(\br+L\hat{\bj},z)$, where $T$ and $L$ are the respective periods. In the paraxial approximation a light beam launched in  such medium is described by the nonlinear Schr\"odinger (NLS) equation for the dimensionless field amplitude $\psi$:
\begin{equation} 
\label{NLS_dimensionless}
    i
    %\frac{\partial\psi}{\partial z}
    \partial_z\psi= H_0(\br,z)\psi-|\psi|^2\psi.  
   % i \frac{\partial \psi}{\partial z} = -\frac{1}{2} \nabla^2 \psi - |\psi|^2\psi + U({\bm r},z) %\psi.
\end{equation}
Here  
\begin{equation}
    \label{H0}
    H_0(\br,z)\equiv -\frac{1}{2}\nabla^2 + U(\br,z),
\end{equation}
is the linear Hamiltonian, $\nabla=(\partial_x,\partial_y)$, and $\br=x\hat{\bi}+y\hat{\bj}$ is the radius-vector in the transverse plane. The last term in Eq.~(\ref{NLS_dimensionless}) accounts for the focusing nonlinearity of the material.
 
\begin{figure*}[ht]
\centering
\includegraphics[width=1\textwidth]{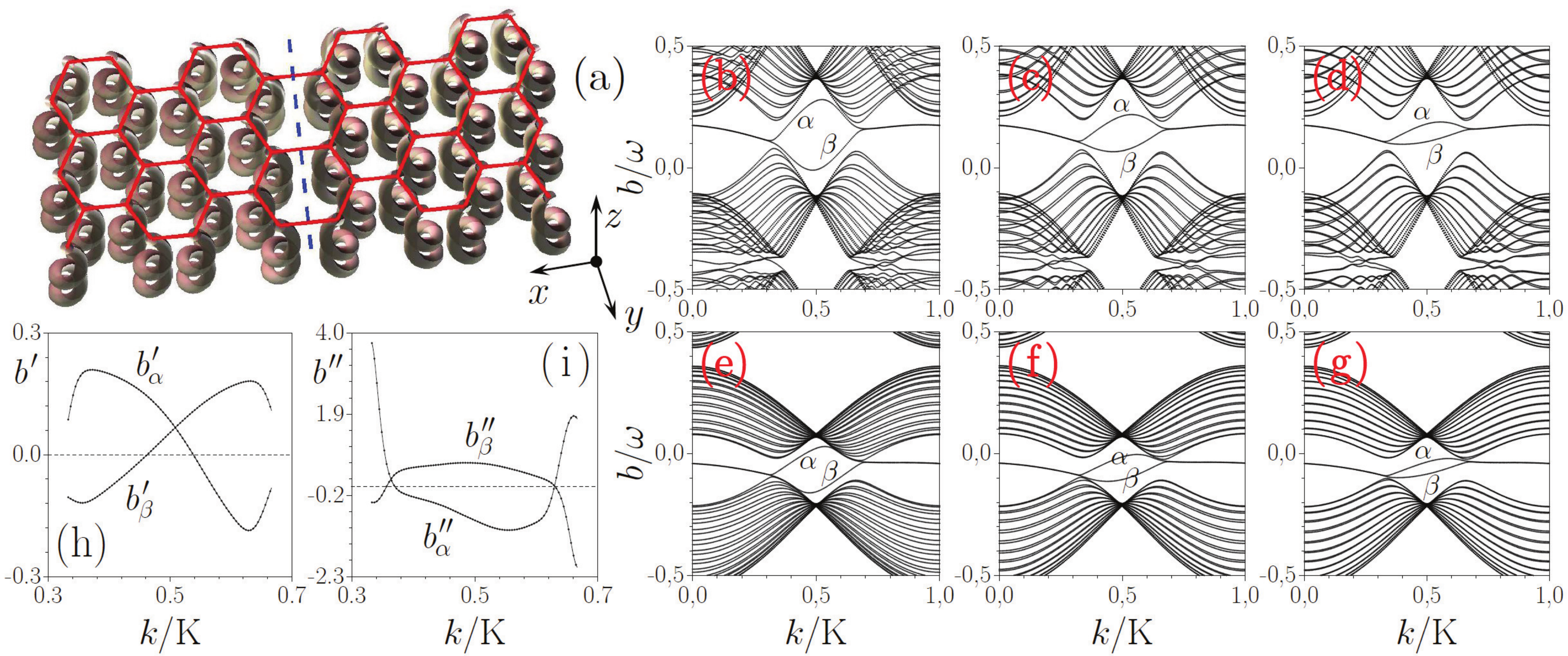}
\caption{Schematic illustration of two arrays of helical waveguides with opposite directions of rotation (a). Propagation constants $b$ of the modes supported by the array versus normalized Bloch momentum $k/K$ for the helix periods $T =10$ (b)-(d) and $T =6$ (e)-(g) and radius $r_0 = 0.4$. Branches marked with $\alpha$ and $\beta$ correspond to the edge states emerging at the zigzag-zigzag interface [see the blue dashed line in (a)]. The spectra are shown for different separations between the arrays: $\delta  = 0.6$ (b),(e), $\delta = 0.9$ (c),(f), and $\delta =1.2$ (d),(g). Negative group velocities $b^\prime$ (h) and group-velocity dispersion $b^{\prime \prime}$ (i) for $\alpha$ and $\beta$ branches of the edge states at $T =10$, $\delta = 0.9$. Dashed lines in (h) and (i) correspond to $b^\prime=0$ or $b^{\prime \prime}=0$.
}
\label{fig:one}
\end{figure*}

\begin{figure}[t]
\centering
\includegraphics[width=0.8\columnwidth]{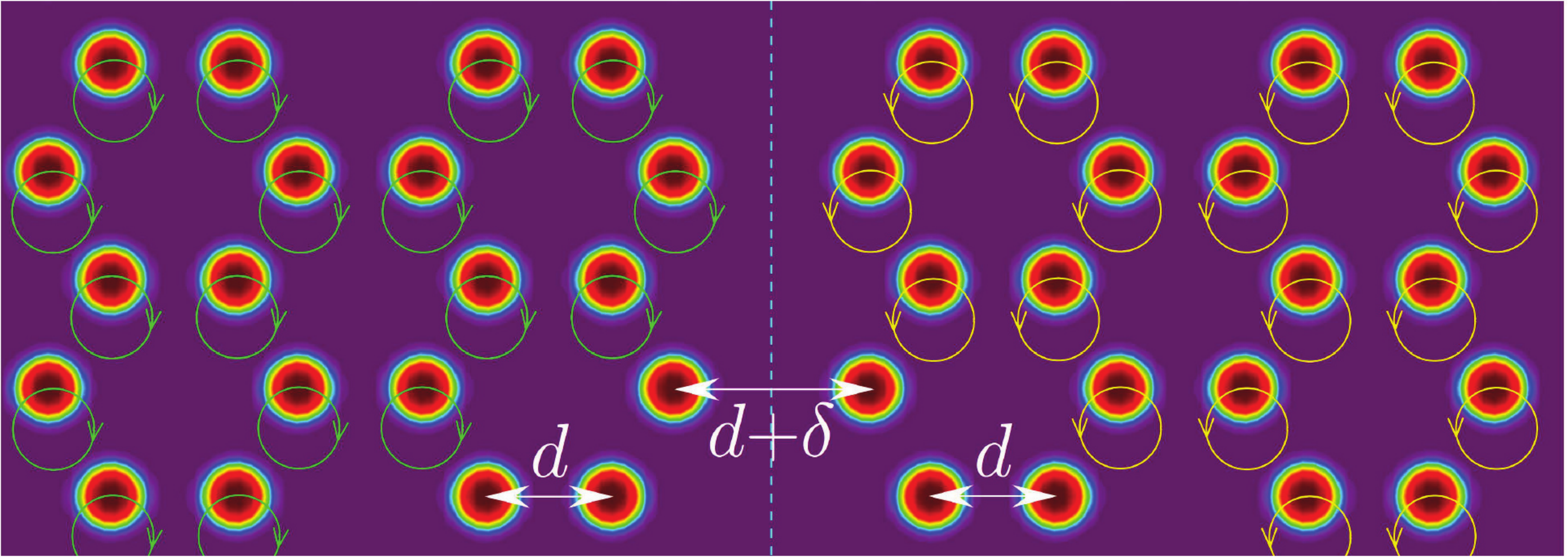}
\caption{
Refractive index distribution in the array of helical waveguides at $z=0$. Green and yellow circles with arrows show waveguide trajectories and rotation directions in $z$. Blue dashed line indicates zigzag-zigzag interface located at $x=0$.}
\label{fig:two}
\end{figure}

An important and experimentally feasible example of such a medium is a combination of two honeycomb arrays of helical waveguides, schematically illustrated in Fig.~\ref{fig:one}(a) (three-dimensional representation) and in Fig.~\ref{fig:two} (two-dimensional cross-section at $z=0$). All waveguides are characterized by equal helix radii $r_0$, but with opposite rotation directions in the two arrays, as shown by the circles with arrows in Fig. \ref{fig:two}. All waveguides have the same longitudinal helix period $T$ along the $z$-axis and are separated by the distance $d$. The first array is located at $-\ell_x\leq x\leq -(\delta+d)/2$ and has a zigzag boundary at $x=-(\delta+d)/2$, while the second array is located at $(\delta+d)/2\leq x\leq \ell_x$ and has zigzag boundary at $x=(\delta+d)/2$. Thus a zigzag-zigzag interface is formed at $x=0$ (it is shown by the dashed blue line in Fig.~\ref{fig:one}(a) and \ref{fig:two}). The introduced parameter $\delta$ is additional spacing along the $x$-axis between two arrays, so that the distance between the nearest waveguides belonging to different arrays at $z=0$ is equal to $d+\delta$ (Fig. \ref{fig:two}). The parameter $\delta$ will be used to control the linear spectrum of the system and it is crucial for the formation of the edge states at the created zigzag-zigzag interface. Both arrays (and hence the entire structure) are periodic along the $y$-axis with a period $L=3^{1/2}d$.

The optical potential describing such arrays can be written for the left $(x<0)$ and right $(x>0)$ arrays marked by the subscripts "$-$" and "$+$", respectively. Evidently, it is $U({\bm r},z)=U_-({\bm r},z)+U_+({\bm r},z)$, where
\begin{equation}
\label{Potential}
         U_\pm({\bm r},z)  = -p\sum_{m,n} e^{
         %\left[
         -\left|{\bm r}-{\bm r}_{nm}\mp \hat{\bi} (\delta+d)/2-r_0 {\bm s}(\pm z)\right|^4/\sigma^4
         %\right]
         }.   
\end{equation}
In this expression $ {\bm s}(z)=\left(\sin(\omega z),\cos(\omega z)\right)$ describes helicity of the waveguides~\cite{bulkFsol,edgeFsol1,discrete2}, ${\bm r}_{mn}$ are the coordinates of the knots of the honeycomb lattice identified by the integers $m$ and $n$, $\sigma$ is the width of a single waveguide, and $p$ is the potential depth. Here we use the dimensionless units in which the transverse coordinates $x,y$ are scaled to the characteristic width $w$, propagation distance $z$ is scaled to the diffraction length $k_0w^2$, where $k_0$ is the wavenumber, while potential depth created by the refractive index modulation is given by $p=\textrm{max}(\delta n)k_0^2w^2/n$. Having in mind a potential realization of our system with fs-laser written waveguide arrays in fused silica, we select dimensionless parameters $d=1.6$, $\sigma = 0.4$, and $p=8.9$ equivalent to $16~\mu\textrm{m}$ separation between neighbouring waveguides of width $4~\mu\textrm{m}$ (hereafter we use characteristic transverse scale $w=10~\mu \textrm{m}$), real refractive index modulation depth of $6.5\cdot 10^{-4}$ at the wavelength $\lambda=800~\textrm{nm}$, and propagation length of $z=1$ corresponding to $1.14~\textrm{mm}$, in accordance with typical experiments \cite{helix1}.

%%%%%%%%%%%%%%%%%%%%%%%%%%%%%%
%%%%%%%%%%%%%%%%%%%%%%%%%%%%%%

Below we consider large, but finite $y$-window: $|y|\leq\ell_y$ with $\ell_y=n L$ where $n\gg 1$ is a large integer, and impose periodic boundary  conditions at $y=\pm\ell_y$. The arrays are considered large enough also along the $x$-axis, i.e., $\ell_x\gg d$. The outer array boundaries in the $x$-direction are bearded [they are not shown in Fig.~\ref{fig:one}(a) and \ref{fig:two}, since we do not use edge states arising at those boundaries]. The boundary conditions at $x=\pm\ell_x$ are formally zero. Since Floquet edge states are exponentially localized around $x=0$ they are not affected by the imposed boundary conditions.

We start with the discussion of a solution $\tpsi$ of the linear problem  
\begin{equation} 
\label{linear}
    i\partial_z\tpsi= H_0(\br,z)\tpsi.  
\end{equation}
%$i\partial_z\psi = H_0\psi$. 
By construction, the  Hamiltonian $H_0$ is invariant with respect to two discrete translations: $H_0(\br+L\hat{\bj},z) =H_0(\br,z+T) =H_0(\br,z)$.  Due to the $T-$periodicity and unitary evolution, one can always define a basis, such that a given state of this basis returns to itself up to some phase~\cite{MoorStad,Rudner-13}. Let us consider one such state (a Floquet state) and denote the respective phase by $\chi$. Such Floquet state can be represented in the form $\tpsi(\br,z)=\phi(\br,z)e^{ibz}$, where $b=\chi/T\in [-\omega/2,+\omega/2)$ with $\omega=2\pi/T$. In this representation $\phi(\br,z)$ is $T-$periodic,  $\phi(\br,z)=\phi(\br,z+T)$, eigenstates of the operator~\cite{well1} $H:=i\partial_z-H_0$,  and $b$ is the corresponding eigenvalue. The operator  $H$ commutes with the $L-$translations along $y-$axis.
Hence at a given $z$ the state $\phi(\br,z)$ has the Bloch-wave form:  $\phi=\phi_{\nu,k}\equiv u_{\nu,k}(\br,z)e^{iky}$, where the functions $u_{\nu,k}$ are both $T$- and $L$-periodic: $u_{\nu,k}(\br,z)=u_{\nu,k}(\br+L\hat{\bj},z)=u_{\nu,k}(\br,z+T)$, $k$ is the Bloch momentum in the first Brillouin zone along the $y$-axis, $k\in[-K/2,+K/2)$ with $K=2\pi/L$, and the index $\nu=1,2,\ldots$ enumerates the allowed bands and the edge states (if any).   
Respectively the Floquet-Bloch states can be  supplied by the band and momentum indexes, and we have:
\begin{equation}
    \label{linear_1}
  H\phi_{\nu,k}=b_{\nu,k}\phi_{\nu,k}.
\end{equation} 
Since, the only change of a state $\psi_{\nu,k}=\phi_{\nu,k}e^{ib_{\nu,k} z}$ over one period $T$ is the phase $\chi_{\nu,k}=b_{\nu,k}T$, we will refer to $b_{\nu,k}$ as propagation constant.

Below we will need only propagation constants for a given momentum $k$, at which nonlinear modes bifurcate from the linear state. For such $k$ the propagation constants can be ordered in accordance with the rule: $b_{\nu_1,k}\geq b_{\nu_2,k}$ for $\nu_2\geq\nu_1$, where the equality takes place if at a given $k$ the level crossing occurs. We will assume that no crossing occurs for topological Floquet-Bloch edge branches.  Floquet-Bloch states can be chosen orthogonal and normalized (see (\ref{normalization}) in Appendix~\ref{app:FB_properties}). 
 
%%%%%%%%%%%%%%%%%%%%%%%%%%%%%%%%%%%%%%%%%%%%%%%%%%%%%%%%%%%%%%%%%
\subsection*{Effective nonlinear Schr\"odinger equation}
\label{subsec:multiscale}
   
Suppose now that the linear array, governed by the Hamiltonian $H_0$, possesses a chiral topological Floquet-Bloch edge state $\psi_{\alpha,k}(\br,z)$ with a given momentum $k$. 
%This edge state is unidirectional (see examples in Sec.~\ref{sec:linear}), since in the system of helical waveguides the time-reversal symmetry is broken. 
We are interested in the family of localized {\em nonlinear} modes bifurcating from the state $\psi_{\alpha,k}(\br,z)$. For  $z$-independent Hamiltonians, families of nonlinear solutions bifurcating from linear modes can be obtained using multiple-scale expansion, whose starting point is the stationary eigenvalue problem for the linear Hamiltonian (see e.g.~\cite{KS,solit-in-latt}). For $z$-dependent Hamiltonians this multiple-scale approach must be considerably revised.

To this end we introduce a formal small parameter $0<\mu\ll 1$, two sets of formally independent scaled variables $(y_0,y_1,y_2,...):=(y, \mu y, \mu^2 y,...)$ and $(z_0,z_1,z_2...):=(z, \mu z, \mu^2 z,...)$, and look for a solution of the NLS equation (\ref{NLS_dimensionless}) with the linear Hamiltonian (\ref{H0}), in the form of the expansion 
\begin{equation}
\label{psi_expan}
\psi_\alpha= \mu e^{ib_{\alpha,k} z_0}\left[A_{\alpha}(y_1,z_1)\phi_{\alpha, k}+ \mu \phi^{(1)}  +\mu^2 \phi^{(2)}+\cdots\right].
\end{equation}
Hereafter we use the convention that in the arguments of the amplitudes only the slowest variables are indicated, e.g. $A_{\alpha}(y_1,z_1)$ stands for a slowly varying amplitude that may depend on the variables $\{y_1,y_2,\ldots\}$ and $\{z_1,z_2,\ldots\}$. The state $\phi_{\alpha, k}$ depends only on the "fast" variables $(y_0,z_0)$: it solves Eq.~(\ref{linear_1}). In the scaled variables the linear Hamiltonian can be rewritten as 
\begin{eqnarray}
 H_0=\tH_0+\mu \tH_1+\mu^2 \tH_2+\cdots\,,
\end{eqnarray}
 where $\tH_0$ is given by (\ref{H0}) with $y$ and $z$ replaced by $y_0$ and $z_0$, $\tH_1= -\partial_{y_0y_1}$, and $\tH_2= -\partial_{y_0y_2}-(1/2)\partial^2_{y_1}$. Next, one substitutes the above expansion (\ref{psi_expan}) into Eq.~(\ref{NLS_dimensionless}), separates the terms in different $\mu$-orders, and solves the obtained equations one by one to define the expansion terms $\phi^{(j)}$, as well as the envelope $A_{\alpha}(y_1,z_1)$.

The choice (\ref{psi_expan}) ensures that the equation, obtained in the first order of $\mu$ is satisfied automatically. Turning to the next orders, we look for functions $\phi^{(j)}$ from (\ref{psi_expan}) in the form of  expansions
\begin{equation}
    \label{expan_first}
    \phi^{(j)}(y_1,z_0) =\sum_\nu C_{\nu,k}^{(j)}(y_1,z_0)\phi_{\nu,k}.
\end{equation}  
To guarantee $T$-periodicity of the functions $\phi^{(j)}(y_1,z_0)$ with respect to $z_0$, the expansion coefficients must be also $T$-periodic: $C_{\nu,k}^{(j)}(y_1,z_0)=C_{\nu,k}^{(j)}(y_1,z_0+T)$. 
Two observations about this expansion are in order. 
Like in the standard multiple-scale expansion used for undriven systems~\cite{KS,solit-in-latt}, the orthogonality of the Bloch states with different Bloch momenta allows one to account in (\ref{expan_first}) only for Bloch modes having the same momentum $k$ as the mode in the leading order has. For this reason no summation over momenta is included in (\ref{expan_first}). At the same time, in contrast to the standard multiple-scale expansion, the sum in (\ref{expan_first}) {\em must} include the term with the leading-order state $\phi_{\alpha,k}(y_1,z_0)$, i.e., at $\nu=\alpha$. This is due to the nature of our periodically driven system, where propagation constant $b_{\alpha,k}$ determines the phase increase after a period $T$ of evolution, rather than at any $z$.

In the $\mu^2-$order we obtain the equation [cf. (\ref{order1})]
\begin{eqnarray}
  \label{mu2order_1}
  i\frac{\partial A_{\alpha}}{\partial z_1}\phi_{\alpha,k}+\frac{\partial A_{\alpha}}{\partial y_1}\frac{\partial \phi_{\alpha ,k}}{\partial y_0} 
\nonumber \\
  =\sum_\nu\left[\frac 1i \frac{  \partial C_{\nu,k}^{(1)} }{\partial z_0}  + (b_{\alpha,k}-b_{\nu,k})C_{\nu,k}^{(1)}\right]\phi_{\nu,k}.
\end{eqnarray}
Projecting this equation on $\phi_{\alpha,k}$ and, using $T$-periodicity of the modes $\phi_{\nu,k}$, and perform averaging over the period $T$ according to the definition
$
\langle f\rangle_T=({1}/{T})\int_{0}^{T}f(\br,z)dz.
$
we arrive at the equation $\partial_{z_1} A_{\alpha} +v_{\alpha,k}\partial_{y_1} A_{\alpha} =0$, in which
\begin{equation}
\label{omega1}
v_{\alpha,k}=-b_{\alpha,k}^\prime =- \left\langle \left(\phi_{\alpha,k},i\partial_ {y_0}\phi_{\alpha,k}\right)\right\rangle_T 
\end{equation}
is the average group velocity of the mode (see Appendix~\ref{kp_method}), and prime stands for derivatives with respect to $k$, e.g. $b_{\alpha,k}^\prime= db_{\alpha,k}/dk$. Thus, the envelope $A_{\alpha}$ depends on the $z_1$ and $y_1$ coordinates only through their combination $Y=y_1-v_{\alpha,k} z_1$, that can be expressed as $A_{\alpha}=A_{\alpha}(Y;y_2,z_2)$. The expansion coefficients $C_{\nu,k}^{(1)}$ entering (\ref{expan_first}) are calculated similarly to the coefficients $c_{\nu,k}^{(1)}$ from Appendix~\ref{kp_method} [see Eqs. (\ref{eq:exp_u})--(\ref{Ax2})] that yields $C_{\nu,k}^{(1)}=-ic_{\nu,k}^{(1)}\partial_{y_1}A_\alpha$.
 
Passing to the $\mu^3-$order equation we observe that the linear dispersive terms in it are considerably simplified after application of $\langle (\phi_{\alpha,k},\cdot)\rangle_T$ to it. Using the expression for the group velocity dispersion  (\ref{omega2}) (derived in Appendix~\ref{kp_method},  where $y$ and $z$ must be replaced by $y_0$ and $z_0$) and the above expression for $C_{\nu,k}^{(1)}$ we obtain 
\begin{eqnarray}
\label{dispersion}
\frac{1}{2}\frac{\partial^2 A_\alpha}{\partial y_1^2}+i\left\langle \sum_{\nu\neq \alpha} \mathcal{V}_{\nu,k}\frac{\partial C_{\nu,k}^{(1)} }{\partial y_1} \right\rangle_T
=-\frac{b_{\alpha,k}^{\prime\prime}}{2}\frac{\partial^2 A_\alpha}{\partial Y^2}
\end{eqnarray}
where
\begin{eqnarray}
\label{Vnu}
 \mathcal{V}_{\nu,k}= -(\phi_{\alpha,k},i\partial_{y_0}\phi_{\nu,k}).
\end{eqnarray}
Finally, we take into account the nonlinear term and require that the envelope $A_\alpha$ is independent of slow variable $y_2$, and employ the Fredholm alternative (i.e., eliminate secular terms). In this way we obtain the effective NLS equation:
\begin{eqnarray}
\label{NLS_main}
i\frac{\partial A_\alpha}{\partial z}-\frac{b_{\alpha,k}^{\prime\prime}}{2}\frac{\partial^2 A_\alpha}{\partial Y^2}+\chi_\alpha |A_\alpha|^2A_\alpha=0,
\end{eqnarray}
where the formal small parameter $\mu$ is set to one, i.e., now $Y=y-v_{\alpha,k}z$,  and the averaged nonlinear coefficient is given by
\begin{equation}
    \label{nonlinearity}
    \chi_\alpha=\langle (|\phi_{\alpha,k}|^2,|\phi_{\alpha,k}|^2)\rangle_T.
\end{equation}

%%%%%%%%%%%%%%%%%%%%%%%%%%%%%%%%%%%%%%%%%%%%%%%%%%%%%%% 
\section*{Spectrum of the linear array}
\label{sec:linear}

To illustrate the possibility of the formation of the edge solitons in our system, we first analyse its Floquet spectrum in the linear limit and establish the existence of topological edge states. To compute the linear spectrum $b_{\alpha,k}$ we use propagation-projection method developed in \cite{cones1}. First, we obtain Bloch modes $\tilde{\phi}_{\nu,k}(\textbf{r})=\tilde{u}_{\nu,k}(\textbf{r})e^{iky}$, where $\tilde{u}_{\nu,k}(\textbf{r})=\tilde{u}_{\nu,k}(\textbf{r}+\hat{{\bm j}}L)$, for the two upper bands in the spectrum of the Hamiltonian $H_0(\br,0)$ defined by (\ref{Potential}) at $z=0$ (it can be interpreted as the Hamiltonian describing the array of straight waveguides). Each such mode $\tilde{\phi}_{\nu,k}$ is normalized according to (\ref{normalization}). Next, these Bloch modes are allowed to evolve, according to (\ref{linear}), along one period $T$ of the linear helical array with a given $r_0>0$. The output distributions $\psi_{\nu,k}^{(T)}(\br)=\tilde{\psi}_{\nu,k}(\br,T)$ corresponding to the inputs $\tilde{\psi}_{\nu,k}(\br,0)=\tphi_{\nu,k}(\br)$, are projected on the input modes that yields $2n\times2n$ monodromy matrix $H_{\nu\nu'}(k,k')=(\tphi_{\nu,k},\psi_{\nu',k'}^{(T)})$ (here $n$ is the number of waveguides in one $y$-period of the structure). The eigenvalues of $H_{\nu\nu'}(k,k')$ obtained in the form of characteristic multipliers $\exp(ib_{\nu,k} T)$ yield the Floquet spectrum $b_{\nu,k}$ of the helical waveguide array. In numerical simulations we found that  $b_{\nu,k}$ feature negligible imaginary parts, for a typical rotation radius $r_0=0.4$ $(4~\mu \textrm{m})$, i.e. radiative losses in helical waveguides are weak.

The obtained linear spectra $b_{\nu,k}$ are illustrated in Figs.~\ref{fig:one}(b)-(g) for two representative helix periods $T=10$ (b)-(d) and $T=6$ (e)-(g), which in the physical units correspond to the periods of $11.39~\textrm{mm}$ and $6.83~\textrm{mm}$. Waveguide rotation opens a topological gap in the spectrum around specific Dirac points $k=K/3$ and $k=2K/3$ \cite{helix1}. Since waveguides in different arrays rotate in the opposite directions stimulating in this way currents in the same direction, \textit{two} branches of the topological edge states appear at the zigzag-zigzag interface between the arrays [shown by the dashed line in Fig.~\ref{fig:one} (a)]. In Fig.~\ref{fig:one} (b)-(g) these branches are denoted by $\alpha$ and $\beta$. The existence of two chiral edge states will be used below for the excitation of topological {\em vector} edge solitons. The topological edge states $\alpha$ and $\beta$ arise in the interval $K/3 < k < 2K/3$ of the Bloch momenta. The edge states emerging in the adjacent intervals $k < K/3$ and $k > 2K/3$ are associated with far-away outer bearded boundaries of two arrays and are not considered here. Notice that the $\alpha$ and $\beta$ edge state branches naturally merge when spacing $\delta$ between two arrays increases and arrays become independent. In contrast, when $\delta \to 0$ the propagation constants of two edge states separate more and more and tend to merge with bulk bands.

Below we typically use helix period $T = 10$, since radiation is negligible in this case. Notice, that at this period one observes Floquet band folding because bands of static array do not fit into longitudinal Brillouin zone of width $2\pi/T$. Further significant increase of the rotation period can lead to stronger folding resulting in closure of the topological gap. The folding disappears for smaller periods. For example, for the period $T = 6$ band folding is not observed since two bands entirely fit into the longitudinal Brillouin zone. At the same time, for this last case in the presence of nonlinearity we will observe the increase of the radiation losses of solitons. Thus, we will need to find the optimal parameters of the arrays such that, on the one hand, the band folding does not close the topological gap, and on the other hand the radiative losses of edge solitons remain negligible.

 Examples of the dependencies of the derivatives $b_{\nu,k}'$ and $b_{\nu,k}^{\prime\prime}$, where $\nu=\alpha,\,\beta$, on $k$ are illustrated in Fig.~\ref{fig:one}(h) and Fig.~\ref{fig:one}(i), respectively. Remarkably, the group velocities, $v_{\nu,k}= -b_{\nu,k}'$, and group velocity dispersion, $v_{\nu,k}'= -b_{\nu,k}''$, of two edge states can have the same as well as opposite signs depending on the $k$ values. The central finding for the existence of \textit{vector} edge solitons (see below) is that there exists the value of Bloch momentum $k\approx 0.51K$ for which group velocities of two edge states are equal, while the respective group velocity dispersion has opposite signs.

%%%%%%%%%%%%%%%%%%%%%%%%%%%%%%%%%%%%%%%%%%%%%%%%%%%%%% 

\section*{Scalar bright and dark edge solitons} \label{sec:dynam}

We now consider \textit{scalar} nonlinear edge solitons. Since $\chi_\alpha>0$, Eq.~(\ref{NLS_main}) predicts a possibility of formation of either bright (if $b''_{\alpha,k}<0$) or dark (if $b''_{\alpha,k}>0$) envelope edge solitons, which bifurcate from the linear Floquet-Bloch edge state $\psi_{\alpha,k}$ belonging to either $\alpha$ or $\beta$ branches. The respective stationary solutions of Eq.~(\ref{NLS_main}) are given by 
\begin{equation} 
\label{Bright_Soliton_Envelope}
    A_{\alpha}= (2b_{\alpha}^{nl}/\chi_\alpha)^{1/2} \; \mathrm{sech}[ (-2b_{\alpha}^{nl}/b''_{\alpha,k})^{1/2}  Y ]e^{-ib_{\alpha}^{nl}z},
\end{equation}
for a bright soliton ($b''_{\alpha,k}<0$) and
\begin{equation}
\label{Dark_Soliton_Envelope}
     A_{\beta}= (b_{\beta}^{nl}/\chi_\beta)^{1/2} \; \tanh[(b_{\beta}^{nl}/b''_{\beta,k})^{1/2}  Y]e^{-ib_{\alpha}^{nl}z},
\end{equation}
for a dark soliton ($b''_{\beta,k}>0$). In these formulas $b_{\alpha}^{nl}>0$ and $b_{\beta}^{nl}>0$ are detunings of the propagation constants from their linear values,  $b_{\alpha,k}$ and $b_{\beta,k}$, arising due to the nonlinearity. They determine also the amplitudes and widths of the solitons.  

\begin{figure*}[ht]
\centering
\includegraphics[width=0.8\textwidth]{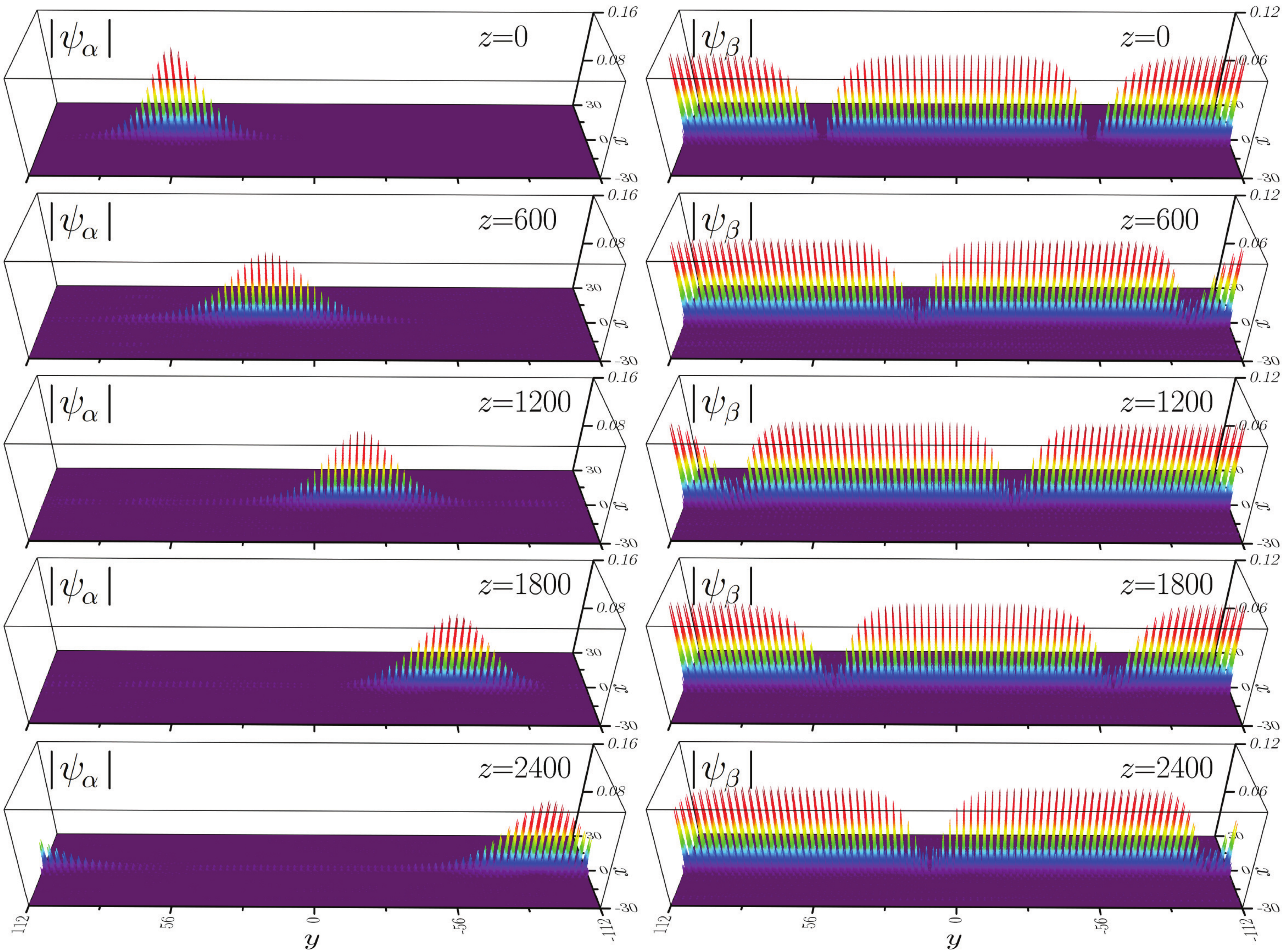}
\caption{Evolution of a \textit{scalar} bright (left column) and \textit{scalar} dark (right column) edge solitons in the array with the helix period $T =10$ and radius $r_0 =0.4$. The solitons belong to the families bifurcating from the linear Floquet modes, $\psi_{\alpha,k}$ and $\psi_{\beta,k}$ with $k =0.51K$, respectively, and correspond to $b_\alpha^{nl} = 0.005$ ($b_\alpha^{\prime\prime} = -0.9143$) and $b_\beta^{nl} = 0.005$ ($b_\beta^{\prime\prime} = 0.6266$). The effective nonlinear coefficients are $\chi_\alpha=0.3876$ and $\chi_\beta=0.4091$. To ensure field periodicity on large, but finite $y$-window a pair of well-separated dark solitons is simultaneously nested on the Bloch wave in the right column.
}
\label{fig:three}
\end{figure*}

In Fig.~\ref{fig:three} we illustrate the evolution of bright (the left column) and dark (the right column) \textit{scalar} edge solitons bifurcating, respectively, from the $\alpha$ and $\beta$ branches at the same Bloch momentum $k =0.51K$ [see Fig.~\ref{fig:one}(i)]. The initial conditions in Eq. (\ref{NLS_dimensionless}) for these soliton solutions are taken as $A_\nu(y,0)\phi_{\nu,k}(\br,0)$ with  $A_\alpha$ and $A_\beta$ given by (\ref{Bright_Soliton_Envelope}) and (\ref{Dark_Soliton_Envelope}), respectively. Here and below for illustration of the evolution of edge solitons we use the separation $\delta=0.9$ ($9~\mu \textrm{m}$ in the physical units) between two arrays.

%%%%%%%%%%%%%%%%%%%%%%%%%%%%%%
%%%%%%%%%%%%%%%%%%%%%%%%%%%%%%

\begin{figure}[t]
\centering
\includegraphics[width=1\columnwidth]{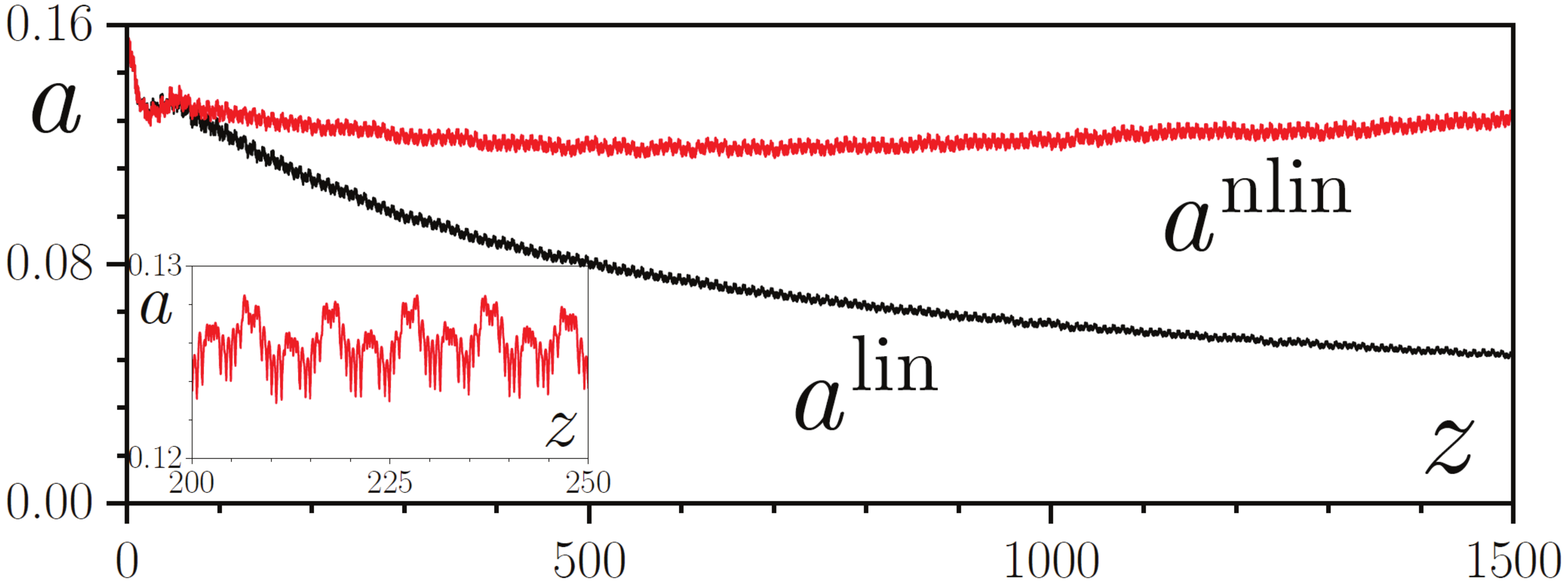}
\caption{The peak amplitude of the scalar bright edge soliton launched into linear and nonlinear medium. All parameters are shown in the caption to Fig.~\ref{fig:two}. The inset shows amplitude evolution over five lattice periods to stress its nearly periodic oscillations.}
\label{fig:four}
\end{figure}

One can see that both bright and dark scalar edge solitons move along the edge without notable modifications in their shapes, even though they traverse more than one hundred of $z$-periods of the structure. There is almost no radiation into the bulk of the lattice for selected $r_0$ and $T$ values. Moreover, no signs of background instability are visible for dark soliton. To prove that these states are indeed supported by the nonlinear self-action in Fig.~\ref{fig:four} we compare the evolution of peak amplitude of the wavepacket (\ref{Bright_Soliton_Envelope}) in nonlinear ($a^{nlin}$, red curve) and linear ($a^{lin}$, black curve) medium. One can see that after the initial transient interval $z\sim 100$, where the peak amplitude $a^{nlin}$ slightly decreases, a soliton acquires an amplitude that on average remains almost constant in $z$. In contrast, without nonlinearity we observe strong asymmetric expansion of the wavepacket and its peak amplitude  $a^{lin}$ substantially decreases.

The existence of the transient period of evolution is a distinctive feature of the excitation of the analytically found envelope edge solitons in a driven system. Such behavior is directly related to other observations as follows. First, the solitons experience small oscillations of their widths (of notches, in the case of dark solitons). Second, the amplitudes of the numerically found solution undergo small $T$-periodic oscillations, as one can see in the inset of Fig.~\ref{fig:four}. In order to understand these phenomena, we recall that
while in the case of gap solitons in stationary arrays, the perturbation arising in the first order of multiple-scale expansion is orthogonal to the wavefunction of the leading order, this is not the case for arrays with helical waveguides. Now the term containing $C_{\alpha,k}^{(1)}(z_0)$ in the expansion (\ref{expan_first}) is nonzero and can be included into the explicit expression for the envelope, thus leading to the correction accounting for oscillations of the soliton velocity. Indeed, let us consider the respective term simultaneously with the leading approximation in the expansion (\ref{psi_expan}). We compute 
\begin{eqnarray}
\label{v_varying}
A_\alpha (y-v_{\alpha,k} z)- c_{\alpha,k}^{(1)}(z) \frac{\partial A_\alpha}{\partial y}\approx A(y-v_{\alpha,k} z- c_{\alpha,k}^{(1)}(z))
\nonumber \\
=A_\alpha \left(y-\int \mathcal{V}_{\alpha,k} (z)dz\right),\qquad
\end{eqnarray} 
where we took into account (\ref{Ax1}).
Thus the maximum of the envelope in fact propagates with the {\em oscillating velocity} $ \mathcal{V}_{\alpha,k}(z)$.  Furthermore, the amplitude $A_\alpha$ found as a result  of averaging over one period, is the envelope of the $T$-periodic mode $\phi_{\alpha, k}(\br,z)$ (see (\ref{psi_expan})). Thus using either (\ref{Bright_Soliton_Envelope}) or (\ref{Dark_Soliton_Envelope}) at $z=0$ we use the initial condition valid for the {\em average} solution in order to simulate dynamics of the {\em exact} solution. On this reason, due to slight reshaping at the initial stage a small fraction of the input power is transferred into radiation.

%%%%%%%%%%%%%%%%%%%%%%%%%%%%%%
%%%%%%%%%%%%%%%%%%%%%%%%%%%%%%

%%%%%%%%%%%%%%%%%%%%%%%%%%%%%%%%%%%%%%%%%%%%%%%%%%%%%%%%%%%%%%%%%
\section*{Vector bright-dark edge solitons}
\label{sec:vector_soliton}

We now turn to the two-component, alias {\em vector}, Floquet edge solitons. Such solitons involve modes belonging to different families of the nonlinear solutions bifurcating from two different branches of the linear Floquet spectrum. A crucial condition for the existence of vector solitons is the equality of the group velocities of the involved edge states for the mentioned momentum $k$. The coupling of the components of vector solitons is mediated by the nonlinearity corresponding to the cross-phase modulation (XPM). Observation of vector edge solitons enabled by periodic driving is not obvious, because envelope equations are obtained only for averaged quantities, while instantaneous group velocity of each component actually oscillates independently, as shown in (\ref{v_varying}). These velocity oscillations are generally different for different modes, i.e., generally speaking $\mathcal{V}_{\alpha,k}(z)\neq \mathcal{V}_{\beta,k}(z)$. In this section we show that in spite of this periodic group-velocity mismatch, vector solitons in Floquet topological insulators do exist.

At the first step, we generalize the average equations for the slowly varying amplitudes to the case of two modes, denoted by subscripts $\alpha$ and $\beta$, which for a given $k$ have equal averaged group velocities, i.e., $v_{\alpha,k}=v_{\beta,k}=v$. In this vector case the envelope-soliton solution is searched in the form [cf. (\ref{psi_expan})]
\begin{eqnarray}
\label{two_psi_expan}
\Psi&=& \mu \left[A_\alpha(y_1,z_1)\psi_{\alpha,k} e^{ib_{\alpha,k} z_0}+   A_\beta(y_1,z_1)\psi_{\beta,k} e^{ib_{\beta,k} z_0}\right]
\nonumber \\  
&+& \sum_{j=1}\mu^{j+1} \left[ \phi_{\alpha,k}^{(j)}e^{ib_{\alpha,k} z_0} +\phi_{\beta,k}^{(j)}e^{ib_{\beta,k} z_0}\right].
\end{eqnarray}
The derivation closely follows the steps outlined for the scalar solitons. Since the effect of the nonlinearity, and hence of the coupling, appears only in the third order of the multiple-scale expansion, we do not show all details of the derivation and present only the final effective coupled NLS equations for the envelopes:
\begin{eqnarray}
\label{two_envel_1}
i\frac{\partial A_\alpha}{\partial z}=\frac{b_{\alpha,k}^{\prime\prime}}{2}\frac{\partial^2 A_\alpha}{\partial Y^2}-\left(\chi_\alpha |A_\alpha|^2+2\chi  |A_\beta|^2\right)A_\alpha,
\\
\label{two_envel_2}
i\frac{\partial A_\beta}{\partial z}=\frac{b_{\beta,k}^{\prime\prime}}{2}\frac{\partial^2 A_\beta}{\partial Y^2}-\left(2\chi |A_\alpha|^2+\chi_\beta  |A_\beta|^2\right)A_\beta,
\end{eqnarray}
where $Y=y-vz$, the nonlinear coefficients $\chi_{\alpha}$, $\chi_{\beta}$ describing self-phase modulation are given by (\ref{nonlinearity}), and the XPM is characterized by $ 
\chi=\langle (|\phi_{\alpha,k}|^2,|\phi_{\beta,k}|^2)\rangle_T
$. 

\begin{figure}[t]
\centering
\includegraphics[width=1\columnwidth]{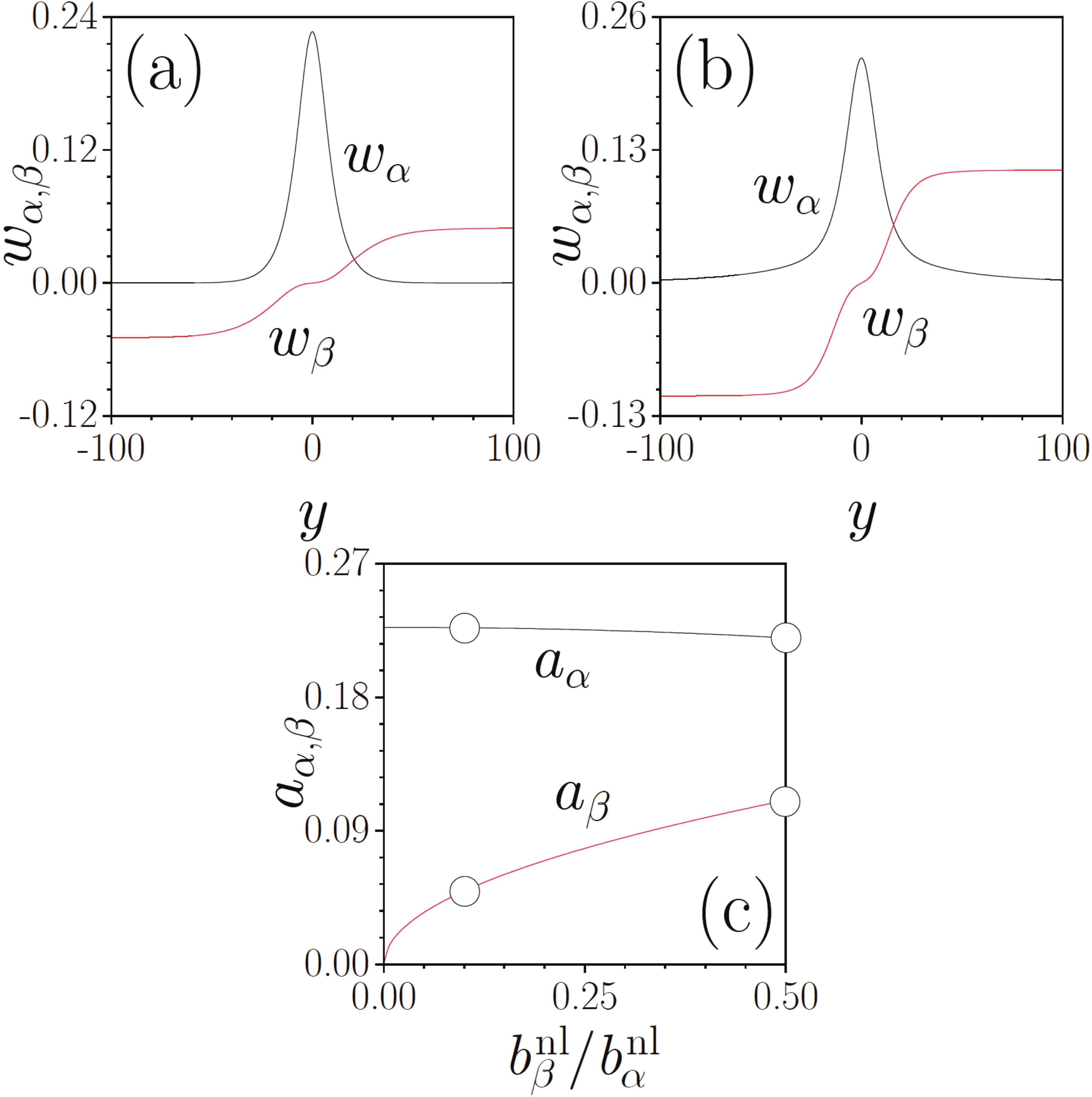}
\caption{
Envelopes of vector edge solitons corresponding to $b_{\beta}^{nl} =0.001$ (a) and $b_{\beta}^{nl} =0.005$ (b) [see dots in panel (c)] and obtained for $b_{\alpha}^{nl} =0.01$, $k =0.51K$. (c) Amplitude of the bright soliton component $a_\alpha=\textrm{max}|u_{\alpha}|$ and of the background of the dark soliton component $a_\beta=\textrm{max}|w_{\beta}|$  of the vector edge solitons {\it versus} $b_{\beta}^{nl}/b_{\alpha}^{nl}$ at $b_{\alpha}^{nl} =0.01$.}
\label{fig:five}
\end{figure}

For the branches $\alpha$ and $\beta$ equal group velocities [crossing of the lines in Fig.~\ref{fig:one}(h)] are achieved at $k=0.51K$, where $4\chi^2-\chi_\alpha \chi_\beta>0$. Taking this into account, in order to ensure the stability of the nonzero background of the mode supporting dark soliton, we consider nonlinear coupling between bright solitons emerging from the branch $\alpha$ and dark solitons emerging from the branch $\beta$. Vector bright-dark soliton solutions of Eqs. (\ref{two_envel_1}), (\ref{two_envel_2}), are found numerically in the form $A_{\nu}=w_{\nu}(Y)\textrm{exp}(ib_{\nu}^{nl}z)$ ($\nu=\alpha,\beta)$ using the Newton method. Examples of the profiles of such vector solitons are presented in Figs.~\ref{fig:five}(a) and (b). We characterize these nonlinear states using the peak amplitudes of the bright $a_\alpha = \max|w_\alpha|$ soliton and  the amplitude of the carrier wave background for the dark soliton $a_\beta = \max|w_\beta|$.   Families of the vector solitons are illustrated in Fig.~\ref{fig:five}(c) by the dependencies of peak amplitudes on the ratio $b_\beta^{nl}/b_\alpha^{nl}$. Solitons exist for $b_\beta^{nl}$ values ranging from $0$ to $b_\alpha^{nl}/2$. At $b_\beta^{nl}\to 0$ the dark component disappears, while at $b_\beta^{nl}\to b_\alpha^{nl}/2$ the bright component looses exponential localization.

%%%%%%%%%%%%%%%%%%%%%%%%%%%%%%
%%%%%%%%%%%%%%%%%%%%%%%%%%%%%%

\begin{figure*}[ht]
\centering
\includegraphics[width=0.8\textwidth]{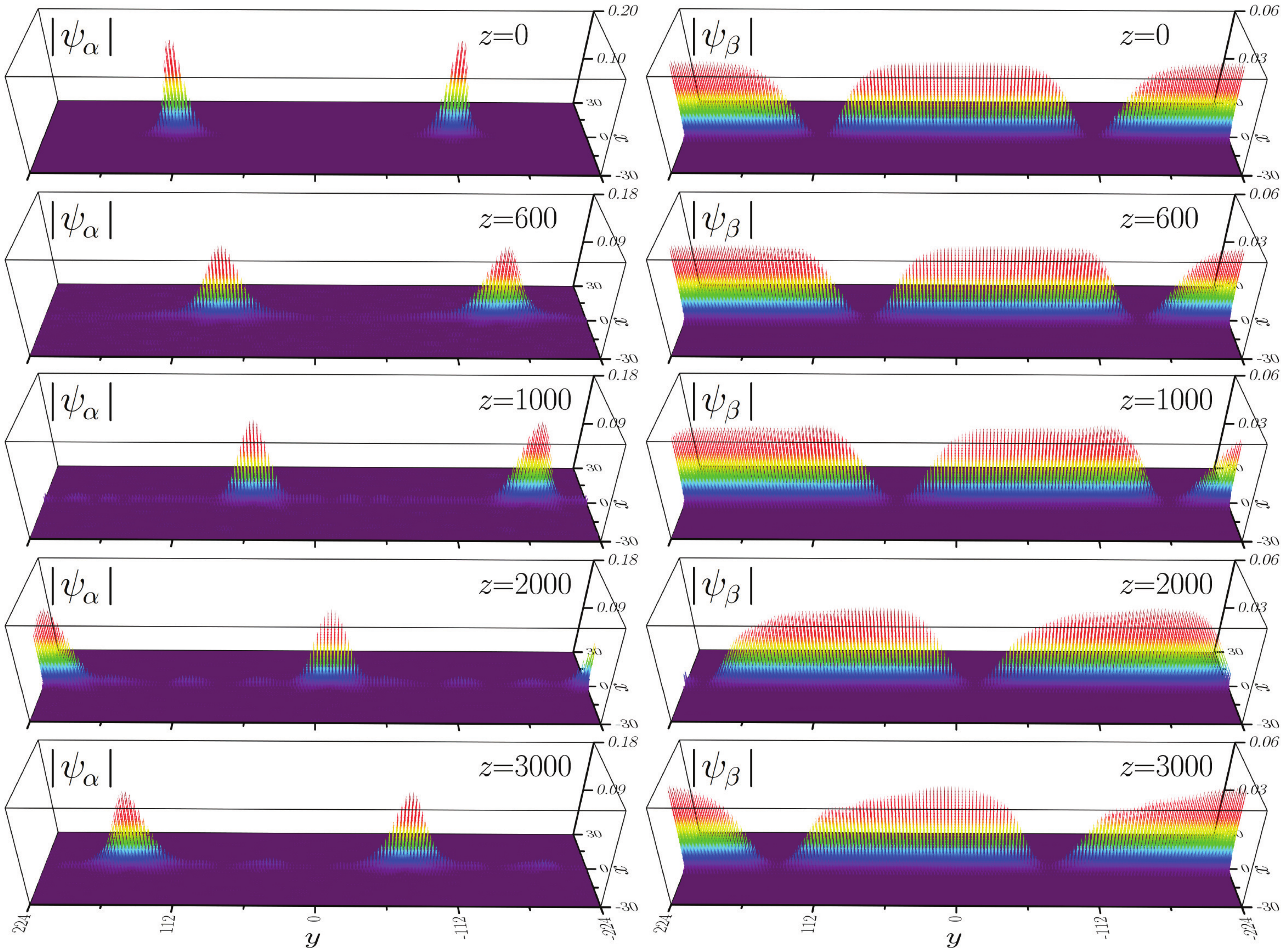}
\caption{Evolution of a \textit{vector} edge soliton with $b_{\alpha}^{nl} =0.010$, $b_{\beta}^{nl} =0.001$ bifurcating from the families of the edge modes  $\alpha$ and $\beta$ at $k =0.51K$. The respective nonlinear coefficients are $\chi_\alpha = 0.3876$, $\chi_\beta = 0.4091$ and $\chi = 0.3965$. The parameters of the array are $T =10$ and $r_0 =0.4$. Bright and dark components are shown in left and right columns, respectively.
%Last row show diffraction in linear medium 
}
\label{fig:six}
\end{figure*}

In Fig.~\ref{fig:six} we show evolution of a dark-bright Floquet edge soliton constructed using numerically obtained envelope from the system (\ref{two_envel_1}), (\ref{two_envel_2}) imposed on the corresponding Bloch modes $\phi_{\alpha,k}$ and $\phi_{\beta,k}$. As in the case of scalar solitons, one can see that at the initial transient stage the peak amplitude of the bright component slightly decreases, after which coupled vector state forms and moves along the interface over considerable distances remaining almost unchanged. Only at the later stages of propagation we observe small distortions of the dark component, presumably arising due to higher-order dispersive effects. The possibility of the formation of vector edge solitons, never discussed before, to the best of our knowledge, is one of the central results of this work.

The theory developed here is accurate for sufficiently broad solitons, whose envelopes cover at least several periods along the interface. Description of  narrower excitations (desirable for experiments) would require inclusion of the higher-order dispersion terms into the equations for the envelopes. We found that radiation from solutions obtained in current second-order approximation notably increases when they shrink to several sites of the structure. However, after initial transient stage, when input narrow soliton broadens, it reaches steady state with new, larger width and propagates as metastable, i.e., very weakly decaying, localized pulse.

%%%%%%%%%%%%%%%%%%%%%%%%%%%%%%
%%%%%%%%%%%%%%%%%%%%%%%%%%%%%%

We have also verified that the results described here remain valid for larger helix radii $r_0$ or smaller helix periods $T$. Increasing $r_0$ or decreasing $T$ is beneficial from the point of view of potential experimental observation of Floquet edge solitons, because this leads to increase of soliton velocity and more pronounced soliton shifts along the edge over experimentally accessible sample lengths. At the same time, this also leads to increasing radiative losses and drives the system into regime, where developed theory may become inapplicable. However, here we propose an approach that still allows one to obtain quasi-solitons even in arrays with smaller helix periods.

\begin{figure}[t]
\centering
\includegraphics[width=0.82\columnwidth]{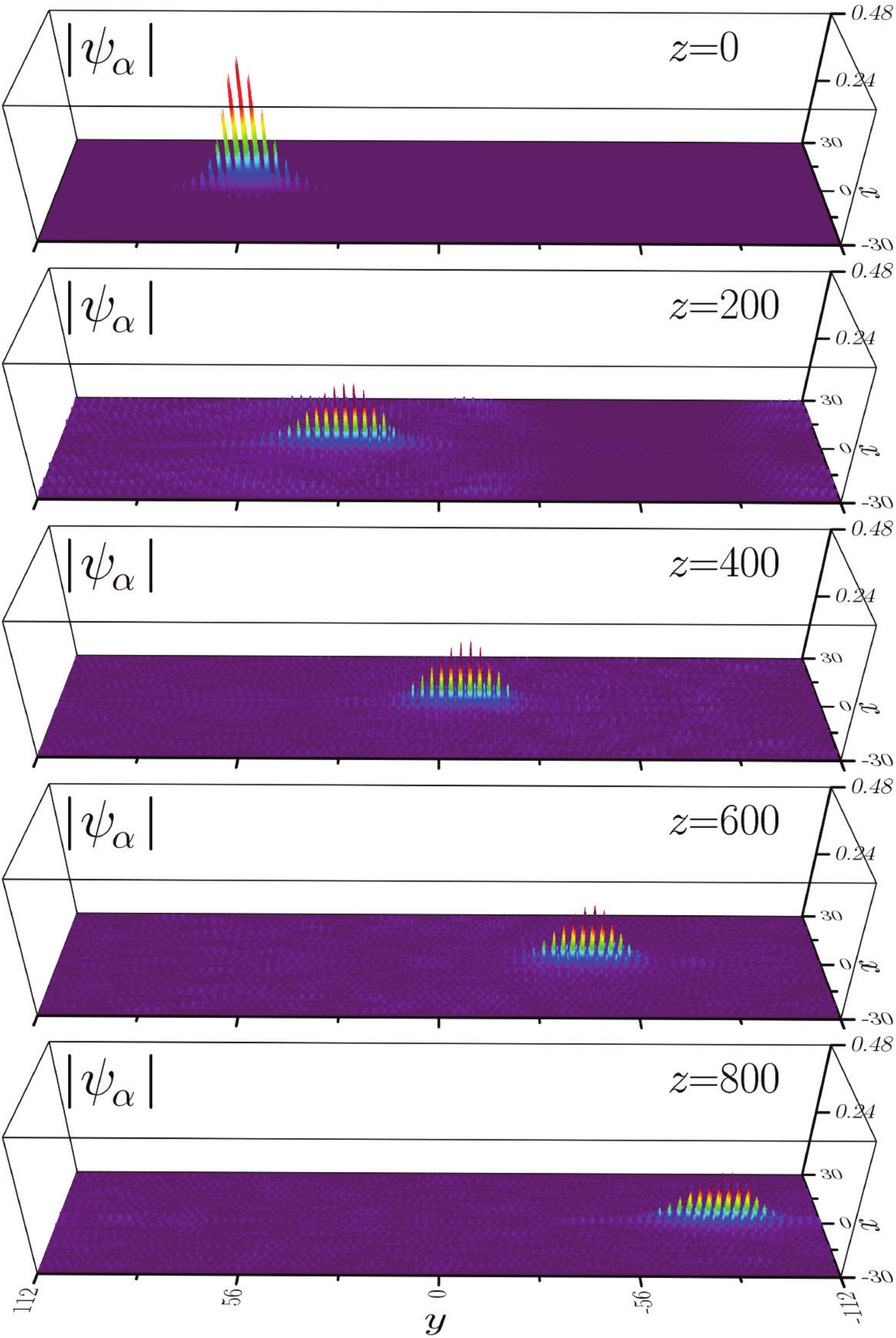}
\caption{Formation of the bright edge soliton at $k =0.51K$ in the array with helix period $T =6$ and radius $r_0 =0.4$. Initial width of the wavepacket corresponds to $b_\alpha^{nl} =0.015$, but its amplitude is 50\% higher than that predicted by Eq.~(\ref{Bright_Soliton_Envelope}). Note substantially larger soliton velocity and radiation in comparison with Fig.~\ref{fig:three}.}
\label{fig:seven}
\end{figure}

For instance, for helix period $T=6$ one can still utilize the formula for the envelope of a bright soliton (\ref{Bright_Soliton_Envelope}), deliberately taking an amplitude somewhat increased in comparison with the analytical prediction (for the same width of the wavepacket). The example of propagation of the scalar bright soliton generated using this kind of the initial conditions is shown in Fig.~\ref{fig:seven}, where the helix radius is $r_0 = 0.4$, the longitudinal period is $T = 6$, and the initial amplitude is increased by $50\%$ in comparison with analytical prediction. Here $b''_\alpha = -0.7944$ and $b_\alpha^{nl} = 0.015$. Even though one can observe a considerable decrease in amplitude at the initial stage of evolution, subsequently the amplitude stabilizes and remains almost constant. Notice substantially larger velocity of soliton motion in comparison with Fig.~\ref{fig:three}. Now one can observe a small amount of radiation into the bulk of the array that gradually reduces the soliton amplitude.

The results presented in Fig. \ref{fig:seven} confirm that topological edge solitons should be observable in fs-laser written helical waveguide arrays. Indeed, already after $z=200$ corresponding to realistic propagation distance of about $22.8~\textrm{cm}$ in accordance with normalization used here, one observes considerable displacement of the edge soliton along the interface that should be easily detectable in the experiment. Moreover, since experiments are typically conducted with finite structures, we verified that our results, obtained for sufficiently large waveguide arrays, remain valid also for finite-width honeycomb ribbons containing only three honeycomb cells at each side of the interface. We found practically no difference in conditions for soliton excitation and in soliton evolution dynamics in such finite structures that can be easily fabricated.

%%%%%%%%%%%%%%%%%%%%%%%%%%%%%%
%%%%%%%%%%%%%%%%%%%%%%%%%%%%%%

\begin{figure}[t]
\centering
\includegraphics[width=1\columnwidth]{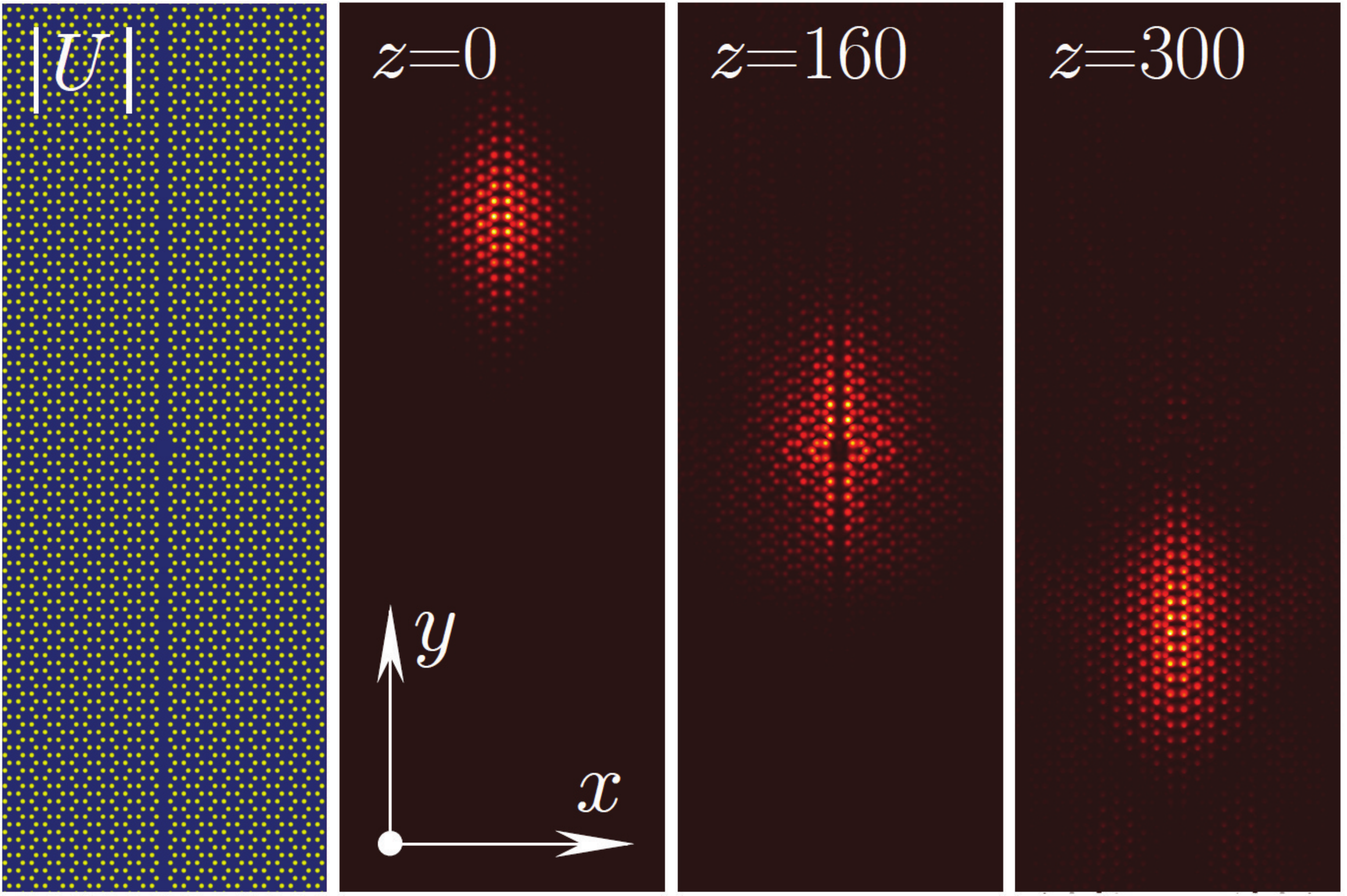}
\caption{Passage of bright edge soliton corresponding to $k =0.39K$, $b_\alpha^{nl} =0.01$ through the defect created by two missing waveguides (see left column) in the array with $T =6$ and $r_0 =0.4$.}
\label{fig:eight}
\end{figure}

Finally, to illustrate that solitons reported here are topologically protected entities that inherit this property from the linear edge states, we consider the interaction of the moving bright edge soliton with a defect in the form of two missing waveguides at the interface between two helical waveguide arrays. Despite the presence of such considerable deformation of the structure, the soliton passes the defect [see Fig.\ref{fig:eight}] practically without radiation into the bulk and without any noticeable backscattering, restoring its shape after the defect and remaining localized. Our estimates show that the passage of the defect by a soliton should be observable experimentally.

%%%%%%%%%%%%%%%%%%%%%%%%%%%%%%
%%%%%%%%%%%%%%%%%%%%%%%%%%%%%%

%%%%%%%%%%%%

\section*{Conclusions}

Summarizing, we have put forward a powerful framework to describe the properties of scalar and vector Floquet edge solitons in a system realized as two helical waveguide arrays with opposite rotation directions of waveguides that form a zigzag-zigzag interface. The theory is based on the multiple-scale expansion with time averaging and it leads to the averaged equations for the envelope of the Floquet-Bloch modes. Based on the developed theory we obtained approximate analytical solutions for stable bright and dark solitons and confirmed their accuracy by direct numerical simulations. The ability of the photonic system we addressed to support two coexisting Floquet-Bloch edge states at one interface  enables the formation of new stable topological objects - vector edge solitons. Remarkably, they are two-dimensional solitons, where confinement in one direction is ensured by the confinement of the Floquet-Bloch mode, while confinement along the interface occurs due to nonlinearity. We verified the accuracy of the developed theory and the stability of the obtained edge solitons using the continuous nonlinear Schr\"odinger equation governing light propagation in media with shallow refractive index modulations. The developed theory illustrates how periodic driving can be used for the design of systems allowing observation of a plethora of new topological nonlinear phenomena. These include the formation of coupled Floquet-Bloch topological soliton states already shown here, as well as new types of self-sustained excitations, such as Bragg topological solitons and topological multi-mode interactions. The presence of the two components in a topological vector solitons is expected to substantially enrich the physics of the soliton interactions. In particular, we conjecture that it opens the possibility to arrange vector edge solitons into stable trains with fixed separation between solitons travelling along the edge, a phenomenon that is not possible with scalar states.

\pagebreak

%%%%%%%%%%%%%%%%%%%%%%%%%%%%%%%%%%%%%%%%%%%%%%%%%%%%%%%%%%%%%%%%%
\begin{acknowledgement}
Y.V.K.\ and L.T.\ acknowledge support from the Severo Ochoa Excellence Programme
%(SEV-2015-0522)
, Fundacio Privada Cellex, Fundacio Privada Mir-Puig, and CERCA/Generalitat de Catalunya. Y.V.K. and S.K.I. acknowledge funding of this study by RFBR and DFG according to the research project numbers 18-502-12080 and SZ 276/19-1. V.V.K. acknowledges support from Portuguese Foundation for Science and Technology (FCT) under Contract no. UIDB/00618/2020. 

\end{acknowledgement}

\appendix
%%%%%%%%%%%%%%%%%%%%%%%%%%%%%%%%%%%%%%%%%%%%%%%%%%%%%%%%%%%%%%%%%
\section{Orthogonality of the Floquet-Bloch states}
\label{app:FB_properties}

Applying $(\phi_{\nu',k'},\cdot) $ where
\begin{equation}
\label{inner}
 (f,g)=\int_{S}f^*(\br,z) g(\br,z)d\br, \quad   S=[-\ell_x,\ell_x]\times [-\ell_y,\ell_y],
\end{equation}
to the relation (\ref{linear_1}), %$H\phi_{\nu,k}=b_{\nu,k}\phi_{\nu,k}$, 
computing its complex conjugate, and using that $b_{\nu,k}$ are real and $H_0$ is Hermitian, one obtains the relation
\begin{equation}
 \label{app1_1}
 i\frac{d}{dz} (\phi_{\nu',k'},\phi_{\nu,k})+(b_{\nu,k}-b_{\nu',k'})(\phi_{\nu',k'},\phi_{\nu,k})= 0.
\end{equation}
Thus, if at any given $z$ two Floquet states are mutually orthogonal, then they are orthogonal at all other $z$. In particular, the orthogonality can be imposed by the choice of the initial conditions $\phi_{\nu,k}(\br,0)=\tilde{\phi}_{\nu,k}(\br)$ at $z=0$, where   $\tilde{\phi}_{\nu,k}(\br)$ are the eigenstates of $H_0(\br,z=0)$, ensuring $(\tphi_{\nu',k'},\tphi_{\nu,k})=\delta_{\nu\nu'}\delta_{kk'}$.  It also follows from (\ref{app1_1}) that $(d/dz) (\phi_{\nu,k},\phi_{\nu,k})=0$. This leads to the normalization condition  
\begin{eqnarray}
 \label{normalization}
 (\phi_{\nu,k}(\br,z),\phi_{\nu',k'}(\br,z))= \delta_{\nu\nu'}\delta_{kk'},
 \end{eqnarray}
where $\delta_{kk'}$ is the Kronecker delta and the  inner product is defined in (\ref{inner}).

%%%%%%%%%%%%%%%%%%%%%%%%%%%%%%%%%%%%%%%%%%%%%%%%%%%%%%%%%%%%%%%%%
\section{${\bm k}\cdot{\bm p}$ method for a periodically driven system}
\label{kp_method}

Here we obtain the direct expressions of the propagation constants, group velocities and group velocity dispersion through the linear Floquet modes. To this end we modify the well-known ${\bm k}\cdot{\bm p}$ method to make it applicable to a periodically driven system. Since now we deal only with the linear problem we do not use the scaled variables; meantime we bear in mind that for applying the obtained results to the multiple scale expansion developed in the text, the variables $y$ and $z$ used in this Appendix must be replaced by $y_0$ and $z_0$, respectively. 

The functions $u_{\nu,k}(x,z)$ solve equation:
\begin{equation}
\label{eq:linear_u}
i\frac{\partial u_{\nu,k}}{\partial z}-b_{\nu,k}u_{\nu,k}= H_ku_{\nu,k}
\end{equation}
where
\begin{equation}
H_k= -\frac{1}{2}\frac{\partial^2}{\partial x^2}+\frac{1}{2}\left(\frac{1}{i}\frac{\partial}{\partial y}+k\right)^2+ U(\br,z) 
\end{equation} 

Consider a state with a band index $\nu=\alpha$ which corresponds to a Bloch wavenumber $k'=k+\delta k$ where $\delta k$ is infinitesimal. On the one hand the Taylor expansion gives
\begin{equation}
\label{eq:om_expan}
b_{\alpha,k+\delta k}=b_{\alpha,k}+b_{\alpha,k}^{\prime}\delta k+\frac{1}{2}b_{\alpha,k}^{\prime\prime}(\delta k)^2+\cdots
\end{equation}
On the other hand equation (\ref{eq:linear_u}) can be rewritten as
\begin{equation}
\label{eq:linear_u_exp}
i\frac{\partial u_{\alpha,k'}}{\partial z}-b_{\alpha,k'}u_{\alpha,k'}= H_ku_{\alpha,k'}+H_1\delta k u_{\alpha,k'}+\frac{1}{2}(\delta k)^2  u_{\alpha,k'}
\end{equation}
where $ 
H_1=-i\partial_y+k
$. 

Now we have to compute $b_{\alpha, k'}$ perturbatively considering $\delta k$ as a small parameter. To this end we use the expansions
\begin{eqnarray}
\label{exp_u_main}
u_{\alpha,k'}=u_{\alpha,k}+\delta k\, u^{(1)}+(\delta k)^2 u^{(2)}+\cdots 
\\
\label{exp_om_main}
b_{\alpha,k^\prime}=b_{\alpha,k}+\delta k\, b^{(1)}+(\delta k)^2 b^{(2)}+\cdots 
\end{eqnarray}
as well as the expansions for $u^{(j)}(\br,z)$ over the subset of functions $u_{\nu,k}(\br,z)$ with the same Bloch vector $k$:  
\begin{eqnarray}
\label{eq:exp_u}
 u^{(j)}(\br,z)=\sum_{\nu} c_{\nu,k}^{(j)}(z) u_{\nu,k} (\br,z)
\end{eqnarray}

All functions $u_{\nu,k}$ are $T-$periodic along $z$. This implies, that all $c_{\nu,k}^{(j)}(z)$ also have to be found as periodic functions of $z$: 
%\begin{eqnarray}
%\label{periodicity}
$
c_{\nu,k}^{(j)}(z+T)=c_{\nu,k}^{(j)}(z).
$
%\end{eqnarray} 

Substituting the above expansions 
%(\ref{exp_u_main}), (\ref{exp_om_main}) and (\ref{eq:exp_u}) 
in Eq.~(\ref{eq:linear_u_exp}), collecting terms up to $(\delta k)^2$ order, we obtain equations in the orders $\delta k$ and $(\delta k)^2$. In the first order we have   
%\begin{widetext}
\begin{eqnarray}
\label{order1}
i\sum_\nu\frac{dc_{\nu,k}^{(1)}}{dz}u_{\nu,k}-\sum_\nu(b_{\alpha,k}-b_{\nu,k})c_{\nu,k}^{(1)}u_{\nu,k} 
\nonumber \\
-b^{(1)} u_{\alpha,k}=H_1u_{\alpha,k}
%\\
%\label{order2}
%i\sum_n\dot{b}_n^{(2)}u_{n,k}-\sum_n(b_{n_0}-b_n)b_n^{(2)}u_{n,k}-\omega^{(1)}\%sum_{n}b_n^{(1)}u_{n,k}- \omega^{(2)}u_{n_0,k}=H_1\sum_n %b_n^{(1)}u_{n,k}+\frac{1}{2}u_{n_0,k}
\end{eqnarray}
%\end{widetext}
%where the overdot stands for the derivative with respect to $z$: $\dot{b}=db/dz$.
Applying $\langle(u_{\alpha,k},\cdot)\rangle_T$ to this equation and using that due to the periodicity $\langle dc_{\alpha,k}^{(j)}/dz\rangle_T=0$,
we compute
 \begin{equation}
\label{omega1_}
b^{(1)}= -\langle  (u_{\alpha,k},H_1u_{\alpha,k})\rangle_T =\langle  (\phi_{\alpha,k},i\partial_{y}\phi_{\alpha,k})\rangle_T.
\end{equation}
Comparing this result with (\ref{eq:om_expan}) we obtain the expression for the average group velocity (\ref{omega1}). 

Next we compute explicit expressions for the coefficients $c_{\nu,k}^{(1)}$ for $\nu=\alpha$
\begin{equation}
\label{Ax1}
c_{\alpha,k}^{(1)}(z) = 
-i\int_{0}^{z}\left[\mathcal{V}_{\alpha,k}(\zeta)-\langle \mathcal{V}_{\alpha,k}\rangle_T \right]   d\zeta,
\end{equation}
and for $\nu\neq\alpha$
\begin{equation}
\label{Ax2}
c_{\nu,k}^{(1)}=\sum_{m=-\infty}^{\infty} \frac{\overline{\mathcal{V}}_{\nu,k}^{m}}{m\omega-b_{\alpha,k}+b_{\nu,k}}e^{- i m\omega z}.
\end{equation}
Here $\mathcal{V}_{\nu,k}^{m}$ are the Fourier coefficients in the expansion of the $T-$periodic function $\mathcal{V}_{\nu,k}(z)$ introduced in (\ref{Vnu}):   
\begin{equation}
\label{h}
\mathcal{V}_{\nu,k} =\sum_{m=-\infty}^{\infty} \mathcal{V}_{\nu,k}^{m} e^{i m \omega z},
\end{equation}
and the overbar stands for the complex conjugation.
Notice that since $|b_{\alpha,k}-b_{\nu,k}|<\omega $ the denominator in (\ref{Ax2}) is different from zero.

Turning to the next order, we apply $\langle(u_{\alpha,k},\cdot)\rangle_T$ to equation obtained in the  $(\delta k)^2$ order. This gives
\begin{eqnarray}
 b^{(2)}=-\frac{1}{2} -\sum_{\nu\neq \alpha}\left\langle (u_{\alpha,k},H_1u_{\nu,k})c_{\nu,k}^{(1)}\right\rangle_T
\end{eqnarray}
where we have taken into account  (\ref{omega1_}). 
Comparing this with (\ref{eq:om_expan}) and returning to the Floquet-Bloch states we obtain the final form of the dispersion of the group velocity in the form
\begin{equation}
\label{omega2}
 b_{\alpha,k}^{\prime\prime}=-1- 2\sum_{\nu\neq \alpha}\left\langle \mathcal{V}_{\nu,k} c_{\nu,k}^{(1)}\right\rangle_T
\end{equation}

%%%%%%%%%%%%%%%%%%%%%%%%%%%%%%%%%%%%%%%%%%%%%%%%%%%%%%%%%%%%%%%%%
\bibliography{VS_bib}

\providecommand{\latin}[1]{#1}
\makeatletter
\providecommand{\doi}
  {\begingroup\let\do\@makeother\dospecials
  \catcode`\{=1 \catcode`\}=2 \doi@aux}
\providecommand{\doi@aux}[1]{\endgroup\texttt{#1}}
\makeatother
\providecommand*\mcitethebibliography{\thebibliography}
\csname @ifundefined\endcsname{endmcitethebibliography}
  {\let\endmcitethebibliography\endthebibliography}{}
\begin{mcitethebibliography}{57}
\providecommand*\natexlab[1]{#1}
\providecommand*\mciteSetBstSublistMode[1]{}
\providecommand*\mciteSetBstMaxWidthForm[2]{}
\providecommand*\mciteBstWouldAddEndPuncttrue
  {\def\EndOfBibitem{\unskip.}}
\providecommand*\mciteBstWouldAddEndPunctfalse
  {\let\EndOfBibitem\relax}
\providecommand*\mciteSetBstMidEndSepPunct[3]{}
\providecommand*\mciteSetBstSublistLabelBeginEnd[3]{}
\providecommand*\EndOfBibitem{}
\mciteSetBstSublistMode{f}
\mciteSetBstMaxWidthForm{subitem}{(\alph{mcitesubitemcount})}
\mciteSetBstSublistLabelBeginEnd
  {\mcitemaxwidthsubitemform\space}
  {\relax}
  {\relax}

\bibitem[Hasan and Kane(2010)Hasan, and Kane]{elect1}
Hasan,~M.~Z.; Kane,~C.~L. Topological insulators. \emph{Rev. Mod. Phys.}
  \textbf{2010}, \emph{82}, 3045\relax
\mciteBstWouldAddEndPuncttrue
\mciteSetBstMidEndSepPunct{\mcitedefaultmidpunct}
{\mcitedefaultendpunct}{\mcitedefaultseppunct}\relax
\EndOfBibitem
\bibitem[Qi and Zhang(2011)Qi, and Zhang]{elect2}
Qi,~X.-L.; Zhang,~S.-C. Topological insulators and superconductors. \emph{Rev.
  Mod. Phys.} \textbf{2011}, \emph{83}, 1057\relax
\mciteBstWouldAddEndPuncttrue
\mciteSetBstMidEndSepPunct{\mcitedefaultmidpunct}
{\mcitedefaultendpunct}{\mcitedefaultseppunct}\relax
\EndOfBibitem
\bibitem[S\"usstrunk and Huber(2015)S\"usstrunk, and Huber]{mech1}
S\"usstrunk,~R.; Huber,~S.~D. Observation of phononic helical edge states in a
  mechanical topological insulator. \emph{Science} \textbf{2015}, \emph{349},
  47--50\relax
\mciteBstWouldAddEndPuncttrue
\mciteSetBstMidEndSepPunct{\mcitedefaultmidpunct}
{\mcitedefaultendpunct}{\mcitedefaultseppunct}\relax
\EndOfBibitem
\bibitem[He \latin{et~al.}(2016)He, Ni, Ge, Sun, Chen, Lu, Liu, and
  Chen]{acou1}
He,~C.; Ni,~X.; Ge,~H.; Sun,~X.-C.; Chen,~Y.-B.; Lu,~M.-H.; Liu,~X.-P.;
  Chen,~Y.-F. Acoustic topological insulator and robust one-way sound
  transport. \emph{Nat. Phys.} \textbf{2016}, \emph{12}, 1124--1129\relax
\mciteBstWouldAddEndPuncttrue
\mciteSetBstMidEndSepPunct{\mcitedefaultmidpunct}
{\mcitedefaultendpunct}{\mcitedefaultseppunct}\relax
\EndOfBibitem
\bibitem[Peng \latin{et~al.}(2016)Peng, Qin, Zhao, Shen, Xu, Bao, Jia, and
  Zhu]{acou2}
Peng,~Y.-G.; Qin,~C.-Z.; Zhao,~D.-G.; Shen,~Y.-X.; Xu,~X.-Y.; Bao,~M.; Jia,~H.;
  Zhu,~X.-F. Experimental demonstration of anomalous Floquet topological
  insulator for sound. \emph{Nat. Commun.} \textbf{2016}, \emph{7}, 13368\relax
\mciteBstWouldAddEndPuncttrue
\mciteSetBstMidEndSepPunct{\mcitedefaultmidpunct}
{\mcitedefaultendpunct}{\mcitedefaultseppunct}\relax
\EndOfBibitem
\bibitem[Goldman \latin{et~al.}(2013)Goldman, Dalibard, Dauphin, Gerbier,
  Lewenstein, Zoller, and Spielman]{cold1}
Goldman,~N.; Dalibard,~J.; Dauphin,~A.; Gerbier,~F.; Lewenstein,~M.;
  Zoller,~P.; Spielman,~I.~B. Direct imaging of topological edge states in
  cold-atom systems. \emph{Proceedings of the National Academy of Sciences of
  the United States of America} \textbf{2013}, \emph{110}, 6736--6741\relax
\mciteBstWouldAddEndPuncttrue
\mciteSetBstMidEndSepPunct{\mcitedefaultmidpunct}
{\mcitedefaultendpunct}{\mcitedefaultseppunct}\relax
\EndOfBibitem
\bibitem[Li \latin{et~al.}(2018)Li, Ye, Chen, Kartashov, Torner, and
  Konotop]{cold2}
Li,~C.; Ye,~F.; Chen,~X.; Kartashov,~Y.~V.; Torner,~L.; Konotop,~V.~V.
  Topological edge states in Rashba-Dresselhaus spin-orbit-coupled atoms in a
  Zeeman lattice. \emph{Phys. Rev. A} \textbf{2018}, \emph{98}, 061601(R)\relax
\mciteBstWouldAddEndPuncttrue
\mciteSetBstMidEndSepPunct{\mcitedefaultmidpunct}
{\mcitedefaultendpunct}{\mcitedefaultseppunct}\relax
\EndOfBibitem
\bibitem[Galilo \latin{et~al.}(2017)Galilo, Lee, and Barnett]{cond1}
Galilo,~B.; Lee,~D. K.~K.; Barnett,~R. Topological Edge-State Manifestation of
  Interacting 2D Condensed Boson-Lattice Systems in a Harmonic Trap.
  \emph{Phys. Rev. Lett.} \textbf{2017}, \emph{119}, 203204\relax
\mciteBstWouldAddEndPuncttrue
\mciteSetBstMidEndSepPunct{\mcitedefaultmidpunct}
{\mcitedefaultendpunct}{\mcitedefaultseppunct}\relax
\EndOfBibitem
\bibitem[St-Jean \latin{et~al.}(2017)St-Jean, Goblot, Galopin, Lema\^{i}tre,
  Ozawa, Gratiet, Sagnes, Bloch, and Amo]{polar1}
St-Jean,~P.; Goblot,~V.; Galopin,~E.; Lema\^{i}tre,~A.; Ozawa,~T.;
  Gratiet,~L.~L.; Sagnes,~I.; Bloch,~J.; Amo,~A. Lasing in topological edge
  states of a 1D lattice. \emph{Nature Photonics} \textbf{2017}, \emph{11},
  651--656\relax
\mciteBstWouldAddEndPuncttrue
\mciteSetBstMidEndSepPunct{\mcitedefaultmidpunct}
{\mcitedefaultendpunct}{\mcitedefaultseppunct}\relax
\EndOfBibitem
\bibitem[Klembt \latin{et~al.}(2018)Klembt, Harder, Egorov, Winkler, Ge,
  Bandres, Emmerling, Worschech, Liew, Segev, Schneider, and
  H\"{o}fling]{polar2}
Klembt,~S.; Harder,~T.~H.; Egorov,~O.~A.; Winkler,~K.; Ge,~R.; Bandres,~M.~A.;
  Emmerling,~M.; Worschech,~L.; Liew,~T. C.~H.; Segev,~M.; Schneider,~C.;
  H\"{o}fling,~S. Exciton-polariton topological insulator. \emph{Nature}
  \textbf{2018}, \emph{562}, 552--556\relax
\mciteBstWouldAddEndPuncttrue
\mciteSetBstMidEndSepPunct{\mcitedefaultmidpunct}
{\mcitedefaultendpunct}{\mcitedefaultseppunct}\relax
\EndOfBibitem
\bibitem[Haldane and Raghu(2008)Haldane, and Raghu]{Haldane}
Haldane,~F. D.~M.; Raghu,~S. Possible Realization of Directional Optical
  Waveguides in Photonic Crystals with Broken Time-Reversal Symmetry.
  \emph{Phys. Rev. Lett.} \textbf{2008}, \emph{100}, 013904\relax
\mciteBstWouldAddEndPuncttrue
\mciteSetBstMidEndSepPunct{\mcitedefaultmidpunct}
{\mcitedefaultendpunct}{\mcitedefaultseppunct}\relax
\EndOfBibitem
\bibitem[Lu \latin{et~al.}(2014)Lu, Joannopoulos, and Solja\v{c}i\'c]{topphot1}
Lu,~L.; Joannopoulos,~J.~D.; Solja\v{c}i\'c,~M. Topological photonics.
  \emph{Nature Photonics} \textbf{2014}, \emph{8}, 821--829\relax
\mciteBstWouldAddEndPuncttrue
\mciteSetBstMidEndSepPunct{\mcitedefaultmidpunct}
{\mcitedefaultendpunct}{\mcitedefaultseppunct}\relax
\EndOfBibitem
\bibitem[Ozawa \latin{et~al.}(2019)Ozawa, Price, Amo, Goldman, Hafezi, Lu,
  Rechtsman, Schuster, Simon, Zilberberg, , and Carusotto]{topphot2}
Ozawa,~T.; Price,~H.~M.; Amo,~A.; Goldman,~N.; Hafezi,~M.; Lu,~L.;
  Rechtsman,~M.~C.; Schuster,~D.; Simon,~J.; Zilberberg,~O.; ; Carusotto,~I.
  Topological photonics. \emph{Rev. Mod. Phys.} \textbf{2019}, \emph{91},
  015006\relax
\mciteBstWouldAddEndPuncttrue
\mciteSetBstMidEndSepPunct{\mcitedefaultmidpunct}
{\mcitedefaultendpunct}{\mcitedefaultseppunct}\relax
\EndOfBibitem
\bibitem[Raghu and Haldane(2008)Raghu, and Haldane]{giro1}
Raghu,~S.; Haldane,~F. D.~M. Analogs of quantum-Hall-effect edge states in
  photonic crystals. \emph{Phys. Rev. A} \textbf{2008}, \emph{78}, 033834\relax
\mciteBstWouldAddEndPuncttrue
\mciteSetBstMidEndSepPunct{\mcitedefaultmidpunct}
{\mcitedefaultendpunct}{\mcitedefaultseppunct}\relax
\EndOfBibitem
\bibitem[Wang \latin{et~al.}(2009)Wang, Chong, Joannopoulos, and
  Solja\v{c}i\'c]{giro2}
Wang,~Z.; Chong,~Y.; Joannopoulos,~J.~D.; Solja\v{c}i\'c,~M. Observation of
  unidirectional backscattering-immune topological electromagnetic states.
  \emph{Nature} \textbf{2009}, \emph{461}, 772--775\relax
\mciteBstWouldAddEndPuncttrue
\mciteSetBstMidEndSepPunct{\mcitedefaultmidpunct}
{\mcitedefaultendpunct}{\mcitedefaultseppunct}\relax
\EndOfBibitem
\bibitem[Hafezi \latin{et~al.}(2011)Hafezi, Demler, Lukin, and Taylor]{res1}
Hafezi,~M.; Demler,~E.~A.; Lukin,~M.~D.; Taylor,~J.~M. Robust optical delay
  lines with topological protection. \emph{Nature Physics} \textbf{2011},
  \emph{7}, 907--912\relax
\mciteBstWouldAddEndPuncttrue
\mciteSetBstMidEndSepPunct{\mcitedefaultmidpunct}
{\mcitedefaultendpunct}{\mcitedefaultseppunct}\relax
\EndOfBibitem
\bibitem[Umucalilar and Carusotto(2012)Umucalilar, and Carusotto]{res2}
Umucalilar,~R.~O.; Carusotto,~I. Fractional Quantum Hall States of Photons in
  an Array of Dissipative Coupled Cavities. \emph{Phys. Rev. Lett.}
  \textbf{2012}, \emph{108}, 206809\relax
\mciteBstWouldAddEndPuncttrue
\mciteSetBstMidEndSepPunct{\mcitedefaultmidpunct}
{\mcitedefaultendpunct}{\mcitedefaultseppunct}\relax
\EndOfBibitem
\bibitem[Rechtsman \latin{et~al.}(2013)Rechtsman, Zeuner, Plotnik, Lumer,
  Podolsky, Dreisow, Nolte, Segev, and Szameit]{helix1}
Rechtsman,~M.~C.; Zeuner,~J.~M.; Plotnik,~Y.; Lumer,~Y.; Podolsky,~D.;
  Dreisow,~F.; Nolte,~S.; Segev,~M.; Szameit,~A. Photonic Floquet topological
  insulators. \emph{Nature} \textbf{2013}, \emph{496}, 196--200\relax
\mciteBstWouldAddEndPuncttrue
\mciteSetBstMidEndSepPunct{\mcitedefaultmidpunct}
{\mcitedefaultendpunct}{\mcitedefaultseppunct}\relax
\EndOfBibitem
\bibitem[Khanikaev \latin{et~al.}(2013)Khanikaev, Mousavi, Tse, Kargarian,
  MacDonald, and Shvets]{meta1}
Khanikaev,~A.~B.; Mousavi,~S.~H.; Tse,~W.-K.; Kargarian,~M.; MacDonald,~A.~H.;
  Shvets,~G. Photonic topological insulators. \emph{Nature Materials}
  \textbf{2013}, \emph{12}, 233--239\relax
\mciteBstWouldAddEndPuncttrue
\mciteSetBstMidEndSepPunct{\mcitedefaultmidpunct}
{\mcitedefaultendpunct}{\mcitedefaultseppunct}\relax
\EndOfBibitem
\bibitem[Bahari \latin{et~al.}(2017)Bahari, Ndao, Vallini, Amili, Fainman, and
  Kant\'e]{laser1}
Bahari,~B.; Ndao,~A.; Vallini,~F.; Amili,~A.~E.; Fainman,~Y.; Kant\'e,~B.
  Nonreciprocal lasing in topological cavities of arbitrary geometries.
  \emph{Science} \textbf{2017}, \emph{358}, 636--640\relax
\mciteBstWouldAddEndPuncttrue
\mciteSetBstMidEndSepPunct{\mcitedefaultmidpunct}
{\mcitedefaultendpunct}{\mcitedefaultseppunct}\relax
\EndOfBibitem
\bibitem[Harari \latin{et~al.}(2018)Harari, Bandres, Lumer, Rechtsman, Chong,
  Khajavikhan, Christodoulides, and Segev.]{laser2}
Harari,~G.; Bandres,~M.~A.; Lumer,~Y.; Rechtsman,~M.~C.; Chong,~Y.~D.;
  Khajavikhan,~M.; Christodoulides,~D.; Segev.,~M. Topological insulator laser:
  Theory. \emph{Science} \textbf{2018}, \emph{359}, eaar4003\relax
\mciteBstWouldAddEndPuncttrue
\mciteSetBstMidEndSepPunct{\mcitedefaultmidpunct}
{\mcitedefaultendpunct}{\mcitedefaultseppunct}\relax
\EndOfBibitem
\bibitem[Bandres \latin{et~al.}(2018)Bandres, Wittek, Harari, Parto, Ren,
  Segev, Christodoulides, and Khajavikhan]{laser3}
Bandres,~M.~A.; Wittek,~S.; Harari,~G.; Parto,~M.; Ren,~J.; Segev,~M.;
  Christodoulides,~D.~N.; Khajavikhan,~M. Topological insulator laser:
  Experiments. \emph{Science} \textbf{2018}, \emph{359}, eaar4005\relax
\mciteBstWouldAddEndPuncttrue
\mciteSetBstMidEndSepPunct{\mcitedefaultmidpunct}
{\mcitedefaultendpunct}{\mcitedefaultseppunct}\relax
\EndOfBibitem
\bibitem[Kartashov and Skryabin(2019)Kartashov, and Skryabin]{laser4}
Kartashov,~Y.~V.; Skryabin,~D.~V. Two-Dimensional Topological Polariton Laser.
  \emph{Phys. Rev. Lett.} \textbf{2019}, \emph{122}, 083902\relax
\mciteBstWouldAddEndPuncttrue
\mciteSetBstMidEndSepPunct{\mcitedefaultmidpunct}
{\mcitedefaultendpunct}{\mcitedefaultseppunct}\relax
\EndOfBibitem
\bibitem[Manakov \latin{et~al.}(1986)Manakov, Ovsiannikov, and
  Rapoport]{pdrive1}
Manakov,~N.~L.; Ovsiannikov,~V.~D.; Rapoport,~L.~P. Atoms in a laser field.
  \emph{Physics Reports} \textbf{1986}, \emph{141}, 320--433\relax
\mciteBstWouldAddEndPuncttrue
\mciteSetBstMidEndSepPunct{\mcitedefaultmidpunct}
{\mcitedefaultendpunct}{\mcitedefaultseppunct}\relax
\EndOfBibitem
\bibitem[Grifoni and H\"{a}nggi(1998)Grifoni, and H\"{a}nggi]{pdrive2}
Grifoni,~M.; H\"{a}nggi,~P. Driven quantum tunneling. \emph{Physics Reports}
  \textbf{1998}, \emph{304}, 229--354\relax
\mciteBstWouldAddEndPuncttrue
\mciteSetBstMidEndSepPunct{\mcitedefaultmidpunct}
{\mcitedefaultendpunct}{\mcitedefaultseppunct}\relax
\EndOfBibitem
\bibitem[Kitagawa \latin{et~al.}(2010)Kitagawa, Berg, Rudner, and
  Demler]{Kitagawa-10}
Kitagawa,~T.; Berg,~E.; Rudner,~M.; Demler,~E. Topological characterization of
  periodically driven quantum systems. \emph{Phys. Rev. B} \textbf{2010},
  \emph{82}, 235114\relax
\mciteBstWouldAddEndPuncttrue
\mciteSetBstMidEndSepPunct{\mcitedefaultmidpunct}
{\mcitedefaultendpunct}{\mcitedefaultseppunct}\relax
\EndOfBibitem
\bibitem[Gu \latin{et~al.}(2011)Gu, Fertig, Arovas, and Auerbach]{Zhenghao-11}
Gu,~Z.; Fertig,~H.~A.; Arovas,~D.~P.; Auerbach,~A. Floquet Spectrum and
  Transport through an Irradiated Graphene Ribbon. \emph{Phys. Rev. Lett.}
  \textbf{2011}, \emph{107}, 216601\relax
\mciteBstWouldAddEndPuncttrue
\mciteSetBstMidEndSepPunct{\mcitedefaultmidpunct}
{\mcitedefaultendpunct}{\mcitedefaultseppunct}\relax
\EndOfBibitem
\bibitem[Lindner \latin{et~al.}(2011)Lindner, Refael, and Galitski]{well1}
Lindner,~N.~H.; Refael,~G.; Galitski,~V. Floquet topological insulator in
  semiconductor quantum wells. \emph{Nature Physics} \textbf{2011}, \emph{7},
  490--495\relax
\mciteBstWouldAddEndPuncttrue
\mciteSetBstMidEndSepPunct{\mcitedefaultmidpunct}
{\mcitedefaultendpunct}{\mcitedefaultseppunct}\relax
\EndOfBibitem
\bibitem[Goldman and Dalibard(2014)Goldman, and Dalibard]{pdrive3}
Goldman,~N.; Dalibard,~J. Periodically Driven Quantum Systems: Effective
  Hamiltonians and Engineered Gauge Fields. \emph{Phys. Rev. X} \textbf{2014},
  \emph{4}, 031027\relax
\mciteBstWouldAddEndPuncttrue
\mciteSetBstMidEndSepPunct{\mcitedefaultmidpunct}
{\mcitedefaultendpunct}{\mcitedefaultseppunct}\relax
\EndOfBibitem
\bibitem[Maczewsky \latin{et~al.}(2017)Maczewsky, Zeuner, Nolte, and
  Szameit]{helix2}
Maczewsky,~L.~J.; Zeuner,~J.~M.; Nolte,~S.; Szameit,~A. Observation of photonic
  anomalous Floquet topological insulators. \emph{Nature Communications}
  \textbf{2017}, \emph{8}, 13756\relax
\mciteBstWouldAddEndPuncttrue
\mciteSetBstMidEndSepPunct{\mcitedefaultmidpunct}
{\mcitedefaultendpunct}{\mcitedefaultseppunct}\relax
\EndOfBibitem
\bibitem[Mukherjee \latin{et~al.}(2017)Mukherjee, Spracklen, Valiente,
  Andersson, \"{O}hberg, Goldman, and Thomson]{helix3}
Mukherjee,~S.; Spracklen,~A.; Valiente,~M.; Andersson,~E.; \"{O}hberg,~P.;
  Goldman,~N.; Thomson,~R.~R. Experimental observation of anomalous topological
  edge modes in a slowly driven photonic lattice. \emph{Nature Communications}
  \textbf{2017}, \emph{8}, 13918\relax
\mciteBstWouldAddEndPuncttrue
\mciteSetBstMidEndSepPunct{\mcitedefaultmidpunct}
{\mcitedefaultendpunct}{\mcitedefaultseppunct}\relax
\EndOfBibitem
\bibitem[Leykam \latin{et~al.}(2016)Leykam, Rechtsman, and Chong]{cones1}
Leykam,~D.; Rechtsman,~M.~C.; Chong,~Y.~D. Anomalous Topological Phases and
  Unpaired Dirac Cones in Photonic Floquet Topological Insulators. \emph{Phys.
  Rev. Lett.} \textbf{2016}, \emph{117}, 013902\relax
\mciteBstWouldAddEndPuncttrue
\mciteSetBstMidEndSepPunct{\mcitedefaultmidpunct}
{\mcitedefaultendpunct}{\mcitedefaultseppunct}\relax
\EndOfBibitem
\bibitem[Noh \latin{et~al.}(2017)Noh, Huang, Leykam, Chong, Chen, and
  Rechtsman]{Noh-17}
Noh,~J.; Huang,~S.; Leykam,~D.; Chong,~Y.~D.; Chen,~K.~P.; Rechtsman,~M.~C.
  Experimental observation of optical Weyl points and Fermi arc-like surface
  states. \emph{Nature Physics} \textbf{2017}, \emph{13}, 611--617\relax
\mciteBstWouldAddEndPuncttrue
\mciteSetBstMidEndSepPunct{\mcitedefaultmidpunct}
{\mcitedefaultendpunct}{\mcitedefaultseppunct}\relax
\EndOfBibitem
\bibitem[Titum \latin{et~al.}(2015)Titum, Lindner, Rechtsman, and
  Refael]{anderson1}
Titum,~P.; Lindner,~N.~H.; Rechtsman,~M.~C.; Refael,~G. Disorder-Induced
  Floquet Topological Insulators. \emph{Phys. Rev. Lett.} \textbf{2015},
  \emph{114}, 056801\relax
\mciteBstWouldAddEndPuncttrue
\mciteSetBstMidEndSepPunct{\mcitedefaultmidpunct}
{\mcitedefaultendpunct}{\mcitedefaultseppunct}\relax
\EndOfBibitem
\bibitem[Titum \latin{et~al.}(2016)Titum, Berg, Rudner, Refael, and
  Lindner]{anderson2}
Titum,~P.; Berg,~E.; Rudner,~M.~S.; Refael,~G.; Lindner,~N.~H. Anomalous
  Floquet-Anderson Insulator as a Nonadiabatic Quantized Charge Pump.
  \emph{Phys. Rev. X} \textbf{2016}, \emph{6}, 021013\relax
\mciteBstWouldAddEndPuncttrue
\mciteSetBstMidEndSepPunct{\mcitedefaultmidpunct}
{\mcitedefaultendpunct}{\mcitedefaultseppunct}\relax
\EndOfBibitem
\bibitem[St\"{u}tzer \latin{et~al.}(2018)St\"{u}tzer, Plotnik, Lumer, Titum,
  Lindner, Segev, Rechtsman, and Szameit]{anderson3}
St\"{u}tzer,~S.; Plotnik,~Y.; Lumer,~Y.; Titum,~P.; Lindner,~N.~H.; Segev,~M.;
  Rechtsman,~M.~C.; Szameit,~A. Photonic topological Anderson insulators.
  \emph{Nature} \textbf{2018}, \emph{560}, 461--465\relax
\mciteBstWouldAddEndPuncttrue
\mciteSetBstMidEndSepPunct{\mcitedefaultmidpunct}
{\mcitedefaultendpunct}{\mcitedefaultseppunct}\relax
\EndOfBibitem
\bibitem[Lustig \latin{et~al.}(2019)Lustig, Weimann, Plotnik, Lumer, Bandres,
  Szameit, and Segev]{Lustig-19}
Lustig,~E.; Weimann,~S.; Plotnik,~Y.; Lumer,~Y.; Bandres,~M.~A.; Szameit,~A.;
  Segev,~M. Photonic topological insulator in synthetic dimensions.
  \emph{Nature} \textbf{2019}, \emph{567}, 356--360\relax
\mciteBstWouldAddEndPuncttrue
\mciteSetBstMidEndSepPunct{\mcitedefaultmidpunct}
{\mcitedefaultendpunct}{\mcitedefaultseppunct}\relax
\EndOfBibitem
\bibitem[Rechtsman \latin{et~al.}(2016)Rechtsman, Lumer, Plotnik, Perez-Leija,
  Szameit, and Segev]{Rechtsman-16}
Rechtsman,~M.~C.; Lumer,~Y.; Plotnik,~Y.; Perez-Leija,~A.; Szameit,~A.;
  Segev,~M. Topological protection of photonic path entanglement. \emph{Optica}
  \textbf{2016}, \emph{3}, 925--930\relax
\mciteBstWouldAddEndPuncttrue
\mciteSetBstMidEndSepPunct{\mcitedefaultmidpunct}
{\mcitedefaultendpunct}{\mcitedefaultseppunct}\relax
\EndOfBibitem
\bibitem[Bandres \latin{et~al.}(2016)Bandres, Rechtsman, and Segev]{Bandres-16}
Bandres,~M.~A.; Rechtsman,~M.~C.; Segev,~M. Topological Photonic Quasicrystals:
  Fractal Topological Spectrum and Protected Transport. \emph{Phys. Rev. X}
  \textbf{2016}, \emph{6}, 011016\relax
\mciteBstWouldAddEndPuncttrue
\mciteSetBstMidEndSepPunct{\mcitedefaultmidpunct}
{\mcitedefaultendpunct}{\mcitedefaultseppunct}\relax
\EndOfBibitem
\bibitem[Lumer \latin{et~al.}(2019)Lumer, Bandres, Heinrich, Maczewsky,
  Herzig-Sheinfux, Szameit, and Segev]{Lumer-19}
Lumer,~Y.; Bandres,~M.~A.; Heinrich,~M.; Maczewsky,~L.~J.; Herzig-Sheinfux,~H.;
  Szameit,~A.; Segev,~M. Light guiding by artificial gauge fields. \emph{Nature
  Photonics} \textbf{2019}, \emph{13}, 339--345\relax
\mciteBstWouldAddEndPuncttrue
\mciteSetBstMidEndSepPunct{\mcitedefaultmidpunct}
{\mcitedefaultendpunct}{\mcitedefaultseppunct}\relax
\EndOfBibitem
\bibitem[Zhang \latin{et~al.}(2018)Zhang, Kartashov, Zhang, Torner, and
  Skryabin]{respol1}
Zhang,~Y.; Kartashov,~Y.~V.; Zhang,~Y.; Torner,~L.; Skryabin,~D.~V. Resonant
  Edge-State Switching in Polariton Topological Insulators. \emph{Las. Photon.
  Rev.} \textbf{2018}, \emph{12}, 1700348\relax
\mciteBstWouldAddEndPuncttrue
\mciteSetBstMidEndSepPunct{\mcitedefaultmidpunct}
{\mcitedefaultendpunct}{\mcitedefaultseppunct}\relax
\EndOfBibitem
\bibitem[Zhang \latin{et~al.}(2018)Zhang, Kartashov, Zhang, Torner, and
  Skryabin]{respol2}
Zhang,~Y.; Kartashov,~Y.~V.; Zhang,~Y.; Torner,~L.; Skryabin,~D.~V. Inhibition
  of tunneling and edge state control in polariton topological insulators.
  \emph{APL Photonics} \textbf{2018}, \emph{3}, 120801\relax
\mciteBstWouldAddEndPuncttrue
\mciteSetBstMidEndSepPunct{\mcitedefaultmidpunct}
{\mcitedefaultendpunct}{\mcitedefaultseppunct}\relax
\EndOfBibitem
\bibitem[Rudner \latin{et~al.}(2013)Rudner, Lindner, Berg, and
  Levin]{Rudner-13}
Rudner,~M.~S.; Lindner,~N.~H.; Berg,~E.; Levin,~M. Anomalous Edge States and
  the Bulk-Edge Correspondence for Periodically Driven Two-Dimensional
  Systemss. \emph{Phys. Rev. X} \textbf{2013}, \emph{3}, 031005\relax
\mciteBstWouldAddEndPuncttrue
\mciteSetBstMidEndSepPunct{\mcitedefaultmidpunct}
{\mcitedefaultendpunct}{\mcitedefaultseppunct}\relax
\EndOfBibitem
\bibitem[Bleu \latin{et~al.}(2016)Bleu, Solnyshkov, and Malpuech]{polsol1}
Bleu,~O.; Solnyshkov,~D.~D.; Malpuech,~G. Interacting quantum fluid in a
  polariton Chern insulator. \emph{Phys. Rev. B} \textbf{2016}, \emph{93},
  085438\relax
\mciteBstWouldAddEndPuncttrue
\mciteSetBstMidEndSepPunct{\mcitedefaultmidpunct}
{\mcitedefaultendpunct}{\mcitedefaultseppunct}\relax
\EndOfBibitem
\bibitem[Kartashov and Skryabin(2016)Kartashov, and Skryabin]{polsol2}
Kartashov,~Y.~V.; Skryabin,~D.~V. Modulational instability and solitary waves
  in polariton topological insulators. \emph{Optica} \textbf{2016}, \emph{3},
  1228--1236\relax
\mciteBstWouldAddEndPuncttrue
\mciteSetBstMidEndSepPunct{\mcitedefaultmidpunct}
{\mcitedefaultendpunct}{\mcitedefaultseppunct}\relax
\EndOfBibitem
\bibitem[Gulevich \latin{et~al.}(2017)Gulevich, Yudin, Skryabin, Iorsh, and
  Shelykh]{polsol3}
Gulevich,~D.~R.; Yudin,~D.; Skryabin,~D.~V.; Iorsh,~I.~V.; Shelykh,~I.~A.
  Exploring nonlinear topological states of matter with exciton-polaritons:
  Edge solitons in kagome lattice. \emph{Scientific Reports volume}
  \textbf{2017}, \emph{7}, 1780\relax
\mciteBstWouldAddEndPuncttrue
\mciteSetBstMidEndSepPunct{\mcitedefaultmidpunct}
{\mcitedefaultendpunct}{\mcitedefaultseppunct}\relax
\EndOfBibitem
\bibitem[Li \latin{et~al.}(2018)Li, Ye, Chen, Kartashov, Ferrando, Torner, and
  Skryabin]{polsol4}
Li,~C.; Ye,~F.; Chen,~X.; Kartashov,~Y.~V.; Ferrando,~A.; Torner,~L.;
  Skryabin,~D.~V. Lieb polariton topological insulators. \emph{Phys. Rev. B}
  \textbf{2018}, \emph{97}, 081103(R)\relax
\mciteBstWouldAddEndPuncttrue
\mciteSetBstMidEndSepPunct{\mcitedefaultmidpunct}
{\mcitedefaultendpunct}{\mcitedefaultseppunct}\relax
\EndOfBibitem
\bibitem[Kartashov and Skryabin(2017)Kartashov, and Skryabin]{polsol5}
Kartashov,~Y.~V.; Skryabin,~D.~V. Bistable Topological Insulator with
  Exciton-Polaritons. \emph{Physical Review Letters} \textbf{2017}, \emph{119},
  253904\relax
\mciteBstWouldAddEndPuncttrue
\mciteSetBstMidEndSepPunct{\mcitedefaultmidpunct}
{\mcitedefaultendpunct}{\mcitedefaultseppunct}\relax
\EndOfBibitem
\bibitem[Lumer \latin{et~al.}(2013)Lumer, Plotnik, Rechtsman, and
  Segev]{bulkFsol}
Lumer,~Y.; Plotnik,~Y.; Rechtsman,~M.~C.; Segev,~M. Self-Localized States in
  Photonic Topological Insulators. \emph{Phys. Rev. Lett.} \textbf{2013},
  \emph{111}, 243905\relax
\mciteBstWouldAddEndPuncttrue
\mciteSetBstMidEndSepPunct{\mcitedefaultmidpunct}
{\mcitedefaultendpunct}{\mcitedefaultseppunct}\relax
\EndOfBibitem
\bibitem[Leykam and Chong(2016)Leykam, and Chong]{edgeFsol1}
Leykam,~D.; Chong,~Y. Edge Solitons in Nonlinear-Photonic Topological
  Insulators. \emph{Phys. Rev. Lett.} \textbf{2016}, \emph{117}, 143901\relax
\mciteBstWouldAddEndPuncttrue
\mciteSetBstMidEndSepPunct{\mcitedefaultmidpunct}
{\mcitedefaultendpunct}{\mcitedefaultseppunct}\relax
\EndOfBibitem
\bibitem[Lumer \latin{et~al.}(2016)Lumer, Rechtsman, Plotnik, and
  Segev]{edgeFsol2}
Lumer,~Y.; Rechtsman,~M.~C.; Plotnik,~Y.; Segev,~M. Instability of bosonic
  topological edge states in the presence of interactions. \emph{Phys. Rev. A}
  \textbf{2016}, \emph{94}, 021801(R)\relax
\mciteBstWouldAddEndPuncttrue
\mciteSetBstMidEndSepPunct{\mcitedefaultmidpunct}
{\mcitedefaultendpunct}{\mcitedefaultseppunct}\relax
\EndOfBibitem
\bibitem[Ablowitz \latin{et~al.}(2014)Ablowitz, Curtis, and Ma]{discrete1}
Ablowitz,~M.~J.; Curtis,~C.~W.; Ma,~Y.-P. Linear and nonlinear traveling edge
  waves in optical honeycomb lattices. \emph{Phys. Rev. A} \textbf{2014},
  \emph{90}, 023813\relax
\mciteBstWouldAddEndPuncttrue
\mciteSetBstMidEndSepPunct{\mcitedefaultmidpunct}
{\mcitedefaultendpunct}{\mcitedefaultseppunct}\relax
\EndOfBibitem
\bibitem[Ablowitz and Cole(2019)Ablowitz, and Cole]{discrete2}
Ablowitz,~M.~J.; Cole,~J.~T. Topological insulators in longitudinally driven
  waveguides: Lieb and kagome lattices. \emph{Phys. Rev. A} \textbf{2019},
  \emph{99}, 033821\relax
\mciteBstWouldAddEndPuncttrue
\mciteSetBstMidEndSepPunct{\mcitedefaultmidpunct}
{\mcitedefaultendpunct}{\mcitedefaultseppunct}\relax
\EndOfBibitem
\bibitem[Konotop and Salerno(2002)Konotop, and Salerno]{KS}
Konotop,~V.~V.; Salerno,~M. Modulational instability in Bose-Einstein
  condensates in optical lattices. \emph{Phys. Rev. A} \textbf{2002},
  \emph{65}, 021602(R)\relax
\mciteBstWouldAddEndPuncttrue
\mciteSetBstMidEndSepPunct{\mcitedefaultmidpunct}
{\mcitedefaultendpunct}{\mcitedefaultseppunct}\relax
\EndOfBibitem
\bibitem[Konotop and Brazhnyi(2004)Konotop, and Brazhnyi]{solit-in-latt}
Konotop,~V.~V.; Brazhnyi,~V.~A. Theory of nonlinear matter waves in optical
  lattices. \emph{Modern Physics Letters B} \textbf{2004}, \emph{18},
  627--651\relax
\mciteBstWouldAddEndPuncttrue
\mciteSetBstMidEndSepPunct{\mcitedefaultmidpunct}
{\mcitedefaultendpunct}{\mcitedefaultseppunct}\relax
\EndOfBibitem
\bibitem[Moore and Stedman(1990)Moore, and Stedman]{MoorStad}
Moore,~D.~J.; Stedman,~G.~E. Non-adiabatic Berry phase for periodic
  Hamiltonians. \emph{J. Phys. A.: Math. Gen.} \textbf{1990}, \emph{23},
  2049--2054\relax
\mciteBstWouldAddEndPuncttrue
\mciteSetBstMidEndSepPunct{\mcitedefaultmidpunct}
{\mcitedefaultendpunct}{\mcitedefaultseppunct}\relax
\EndOfBibitem
\end{mcitethebibliography}

%\newpage
%%%%%%%%%%%%%%%%%%%%%%%%%%%%%%
%\begin{figure*}[ht]
%\centering
%\includegraphics[width=1\textwidth]{For Table of Contents Only.jpg}
%\end{figure*}
%%%%%%%%%%%%%%%%%%%%%%%%%%%%%%

\end{document}